# Using Generative AI to Uncover What Drives Player Enjoyment in PC and VR Games

Hisham Abdelqader[1]

**Abstract**
As video games continue to evolve, understanding what drives player enjoyment remains a key challenge. Player reviews provide valuable insights, but their unstructured nature makes large-scale analysis difficult. This study applies generative AI and machine learning, leveraging Microsoft Phi-4 small language model (SLM) and Google Cloud, to quantify and analyze game reviews from Steam and Meta Quest stores. The approach converts qualitative feedback into structured data, enabling comprehensive evaluation of key game design elements, monetization models, and platform-specific trends. The findings reveal distinct patterns in player preferences across PC and VR games, highlighting factors that contribute to higher player enjoyment. By using Google Cloud for large-scale data storage and processing, this study establishes a scalable framework for game review analysis. The study's insights offer actionable guidance for game developers, helping optimize game mechanics, pricing strategies, and player engagement.

## 1 Introduction

The global video game market size is projected to grow from USD 248.52 billion in 2023 to USD 664.96 billion by 2033, according to GlobeNewswire, registering a Compound Annual Growth Rate (CAGR) of 10.32% [1]. This indicates that the video game industry is not only expanding rapidly but also has a substantial impact on the global economy.

As the video game industry continues its rapid expansion, game reviews become essential as they assess the enjoyment factor in games, which is the core element that keeps players engaged. Reviews help identify both the strengths and weaknesses of a game, providing valuable feedback for developers and players alike. They also play a crucial role in determining a game's success or failure, influencing players' decisions on whether to invest their time and money in a particular title. Additionally, reviews serve as a tool for understanding what makes a game enjoyable while highlighting areas for potential improvement [2], [3]. However, gamers can be a particularly demanding audience, with many holding high expectations regarding game quality [4].

Existing research primarily focused on analyzing player feedback through text-mining and natural language processing (NLP) techniques only [5], [6], [7]. For example, techniques such as Latent Dirichlet Allocation (LDA), a topic modeling method, and TextBlob, an NLP library, were used to extract valuable insights from player reviews regarding their preferences, challenges, and experiences across various video games [8], [9]. While these techniques are powerful tools for large-scale analysis, they often struggle to capture the more nuanced aspects of player feedback, such as emotional complexity, sarcasm, or cultural context, thereby limiting the depth of the insights generated. As a result, they may fail to accurately interpret emotions of enjoyment toward a particular game or contextual descriptions related to specific game design elements [10], [11].

This study aims to lay the foundation for a deeper understanding of the factors that contribute to player enjoyment in video games across both personal computer (PC) and virtual reality (VR) platforms, as well as various genres. By leveraging generative AI techniques, such as automated text summarization, sentiment analysis, and topic modeling, an empirical analysis will be conducted of player reviews from Steam and Meta Quest stores. This approach enables the identification of key elements like gameplay mechanics, story and characters, level design, and monetization strategies that correlate with positive player experiences. The insights gained from this study can inform game developers and designers about what aspects resonate most with players, guiding the creation of more engaging and enjoyable games.

Additionally, it provides a data-driven basis for future research in game design, player psychology, and user experience, ultimately contributing to advancements in the gaming industry and academic studies related to interactive media. Based on the analyzed literature, the research questions are:

- RQ1: What are the key game design features that contribute most significantly to player enjoyment in PC versus VR games?

Hisham Abdelqader
ha396@uowmail.edu.au

[1] School of Computer Science, University of Wollongong in Dubai, Internet City, UAE

- RQ2: In what ways do monetization strategies affect player enjoyment in free-to-play versus non-free-to-play games?
- RQ3: What differences exist in player enjoyment between PC and VR game genres?
- RQ4: How can generative AI techniques be employed to identify the primary factors influencing positive player reviews across different game genres and platforms?

Given the rapidly evolving nature of the gaming industry, this study focuses on games released between 2020 and 2024, a period marked by significant shifts in player behavior following the COVID-19 pandemic. While this temporal scope allows us to capture emerging trends, it may not fully represent longer-term patterns in gaming preferences. Moreover, user-generated reviews, while rich in insights, introduce challenges such as review bombing, where mass negative reviews may distort sentiment analysis and misrepresent player enjoyment [7]. Finally, while generative AI techniques facilitate large-scale text analysis, they may struggle with linguistic nuances like sarcasm and humor, potentially impacting sentiment classification accuracy [12].

The remainder of this paper is organized as follows. Section 2 reviews prior work on generative AI, machine learning techniques in video game review analysis, and studies on player enjoyment and monetization strategies. Section 3 describes the methodology, including data collection from Steam and Meta Quest reviews and the application of generative AI for text analysis and feature extraction. Section 4 presents the results of the statistical and machine learning analyses. Finally, Sections 5 and 6 discuss the findings, highlight their implications for game developers, outline limitations, and suggest directions for future research.

## 2 Literature review

### 2.1 Popular video games platforms

Valve is a major player, primarily through its Steam platform, which is the largest digital distribution platform for PC games. Steam supports a wide variety of video games, and its compatibility with multiple head-mounted displays (HMDs) makes it a significant player in the personal computer virtual reality (PCVR) space [8]. PCVR refers to virtual reality (VR) experiences that rely on a personal computer for processing power. In this setup, the headset is tethered to the PC via cables and mainly serves as a display and motion-tracking device, while the intensive graphics rendering and computational tasks are performed by the PC's high-end graphics processing unit (GPU) and central processing unit (CPU). This allows PCVR to deliver more immersive, visually detailed, and high-fidelity experiences than standalone headsets, which run applications directly on their built-in hardware [13].

It is important to note that, unlike Meta's store, Steam does not use a numerical rating system for game reviews. Instead, players can rate a game as either 'Recommended' (positive review) or 'Not Recommended' (negative review). In contrast, Meta employs a numerical rating system ranging from 0 to 5, where 0 indicates not recommended at all, while 5 signifies highly recommended. These reviews also include a textual component where players share their personal experiences. While anyone can read these reviews, Valve Corporation provides an application programming interface (API) that enables the scraping of review data. This scraped data can then be analyzed using various methods to understand player experiences [14] or to evaluate the helpfulness of reviews.

The Meta Quest store serves as the official VR app store for Meta's HMDs. It can be accessed both within VR and in 2D on PCs and mobile devices. The platform comprises four subcategories: Quest, Rift, Go, and GearVR. As of July 2022, the store offers a total of 3,547 VR games. Meta boasts a significant user base of over 1,490,000, supported by three primary types of HMDs: phone-powered VR (e.g., GearVR), PC VR (e.g., Rift, Rift S), and all-in-one standalone devices (e.g., Quest, Quest 2, Quest 3, Quest 3s). Notably, the Meta Quest store is exclusive to Meta series HMDs and is not compatible with devices from other manufacturers [15].

Meta dominates the market with around 60% of VR headset distribution (Oculus Rift S, Meta Quest 2, Meta Quest 3) which is illustrated in Fig. 1. Thus, the Meta Quest store is a reliable data source due to its exclusive focus on VR content. Unlike platforms such as SteamVR or PlayStation VR, which support both PC and VR games, the Meta Quest store offers only VR-specific experiences [9].

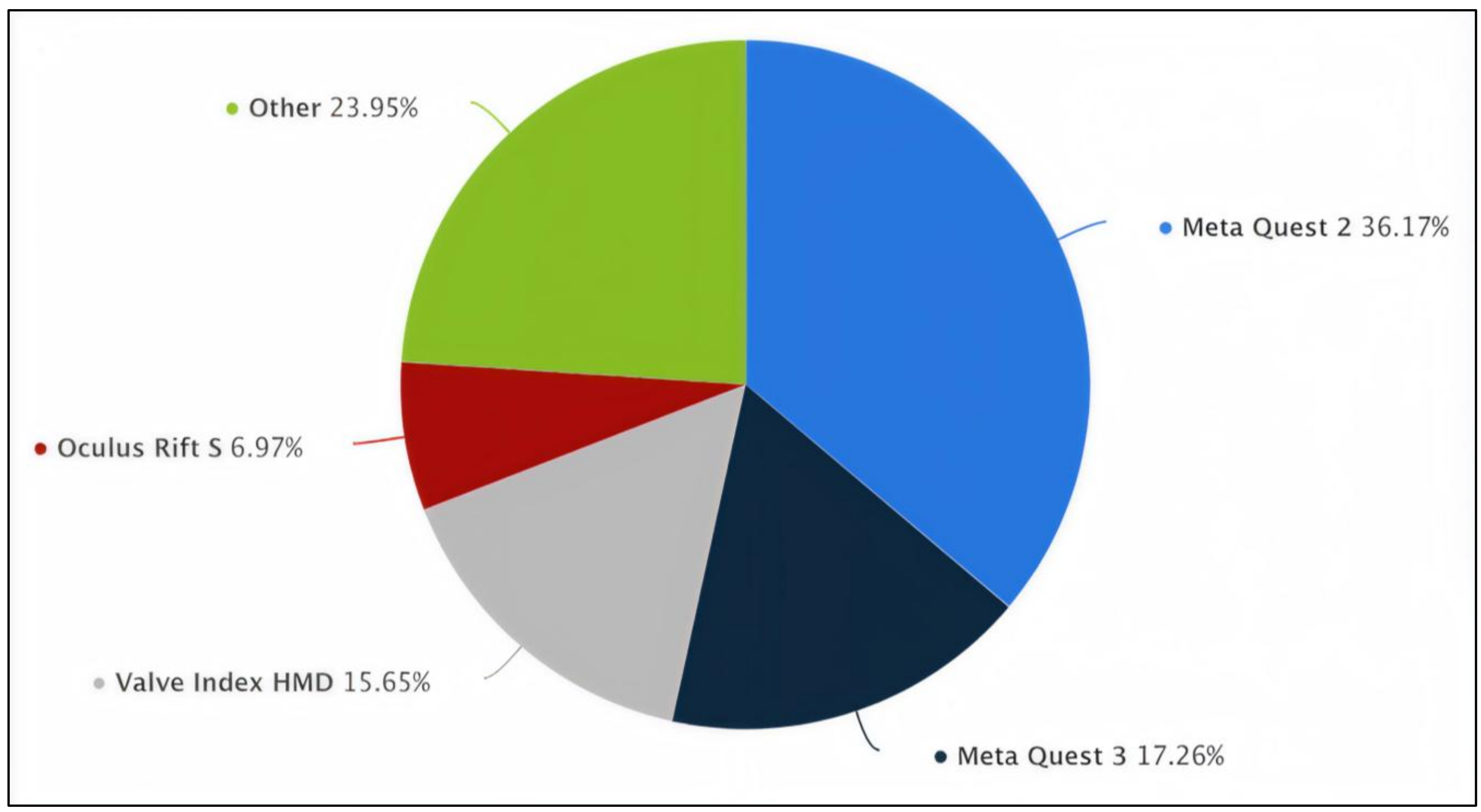

**Fig. 1** Steam users' share of VR headsets by device [16]

### 2.2 Limitations of existing research

Several studies have analyzed reviews from different digital video game distribution platforms. Pagano and Maalej found that most reviews are submitted shortly after an application's release, with the frequency declining rapidly over time. They also noted that reviews often cover multiple topics such as feature requests, user experience, and bug reports [17]. Guzsvinecz focused on PC video game reviews using textual data from Steam. Lin and colleagues analyzed reviews of early access titles and specific top-level genres (TLGs) and genres [6], while Guzsvinecz examined the effects of playtime and game mechanics on reviews within the Souls-like subgenre [5].

While Viggiato and Bezemer in [18] leveraged Transformer models like OPT-175B and Yang Yu et al. in [19] employed Bidirectional Encoder Representations from Transformers (BERT) for sentiment analysis of game reviews, they overlooked several crucial elements that influence player enjoyment, such as gameplay mechanics, story development, level design, and the immersive qualities unique to VR games.

Another limitation is the genre-specific focus of some studies. For instance, Guzsvinecz's focus on "Souls-like" games provided valuable insights into that particular genre like correlations between positive reviews, playtime, and game design features such as graphics and style but limited the applicability of the findings to other types of games [5].

Although monetization strategies are central to the gaming industry's economic model and heavily influence user enjoyment, only Dayi and his team addressed monetization strategies in their analysis of Steam game reviews [6]. Their study merely noted that free-to-play video games tend to receive shorter reviews compared to non-free-to-play games in terms of text length. However, it did not explore the correlation between monetization and player enjoyment.

The focus on analyzing English-language reviews in several studies, including [4], [6], [10], and [19], limits the generalizability of findings by excluding non-English-speaking players and diverse cultural perspectives. This introduces bias by focusing on Western markets, neglecting significant gaming regions such as Asia. Additionally, the linguistic nuances and sentiment expressions unique to non-English reviews are missed, potentially affecting the accuracy of sentiment analysis. As a result, these studies do not fully capture the global gaming community's feedback, limiting the breadth of insights into player experiences.

### 2.3 Summary

The comparative analysis in Table 1 further reinforces the gaps identified in previous research and underscores how our study advances the field. The references to the journals corresponding to Table 1 are: JA1: [4], JA2: [6], JA3: [5], JA4:

[8], JA5: [10], JA6: [20], JA7: [7], JA8: [9], JA9: [18], JA10: [11], JA11: [19]. All these studies were published between 2019-2024.

**Table 1** Similar works comparison table

| Journals | Uses Steam reviews | Uses Meta reviews | Uses generative AI | Correlation with genres | Correlate with design elements | Correlate with monetization | Multilingual analysis of reviews | Number of games analyzed | Number of reviews analyzed |
|---|---|---|---|---|---|---|---|---|---|
| JA1 | Yes | No | No | Yes | No | No | No | NA | 35,983,481 |
| JA2 | Yes | No | No | No | Yes | Yes | No | 6,224 | 12,338,364 |
| JA3 | Yes | No | No | No | Yes | No | No | 21 | 993,932 |
| JA4 | Yes | Yes | No | No | No | No | No | 7,413 | 176,428 |
| JA5 | Yes | No | No | No | Yes | No | No | 1 | 99,993 |
| JA6 | Yes | Yes | No | No | No | No | No | 9 | 105,757 |
| JA7 | Yes | No | No | No | No | No | No | 750 | 17,635 |
| JA8 | No | Yes | No | Yes | Yes | No | No | 157 | 115,531 |
| JA9 | Yes | No | No | No | No | No | No | NA | 614,403 |
| JA10 | Yes | No | Yes | No | No | No | Yes | 6,224 | 12,383,364 |
| JA11 | Yes | No | Yes | Yes | Yes | No | No | 4 | 1,600,000 |
| **Our Study** | **Yes** | **Yes** | **Yes** | **Yes** | **Yes** | **Yes** | **1-11** | **4,856** | **485,600** |

While multiple studies have analyzed Steam game reviews, few have combined this dataset with Meta game reviews, which offer a different perspective on player sentiment. Our study is among the few that integrate both sources, allowing for a more comprehensive analysis that bridges platform-specific biases.

Moreover, while prior studies have explored correlations between reviews and game genres, game mechanics, or monetization strategies, these aspects have largely been examined in isolation. Table 1 highlights that most existing research tends to focus on only one or two of these factors rather than considering them holistically. Our study addresses this limitation by simultaneously examining correlations with game genres, gameplay elements—including story, mechanics, and artistic aspects—and monetization models. This multifaceted approach provides a richer understanding of how various design and economic factors shape player sentiment.

Another important distinction is our study's adoption of generative AI, which remains largely unexplored in existing research. By leveraging generative AI, our study is able to conduct a more nuanced analysis of player feedback, moving beyond sentiment classification to uncover underlying trends and emergent themes in game reviews.

Additionally, Table 1 underscores the significant variation in the number of games and reviews analyzed across studies. While some studies focus on only a handful of titles, others examine tens of thousands of reviews without necessarily offering detailed insights into game design factors. Our study strikes a balance by analyzing a substantial dataset of 4,856 games and 485,600 reviews, ensuring both depth and breadth in our findings.

Finally, the inclusion of multilingual reviews in our study marks a notable departure from prior research, which has predominantly focused on English-language data. As illustrated in Table 1, only a few studies have considered reviews in multiple languages. By addressing this gap, our study provides a more globally representative perspective on player experiences, capturing diverse linguistic and cultural contexts that are often overlooked.

In sum, the comparative findings from the literature review substantiate the need for a more integrative and expansive approach to game review analysis. By leveraging a diverse dataset, incorporating generative AI, and examining multiple interrelated factors, our study offers a more comprehensive and methodologically advanced contribution to the field.

## 3 Methodology

### 3.1 Overview

To answer the RQs, reviews of all video games on the Steam and Meta Quest stores from 2020-2024 were scraped and analyzed. This section is divided into two parts: Subsection 3.2 outlines the scraping process, subsections 3.3 and 3.4 detail the data processing and preparation. Finally, subsection 3.5 describes data visualization. All data analyses and visualizations were conducted using Google Cloud [21]. Scraping code and functions used in the analysis were sourced from Github [22], [23] and PyPi [24]. Figure 2 provides an overview of the methodology adopted in this study.

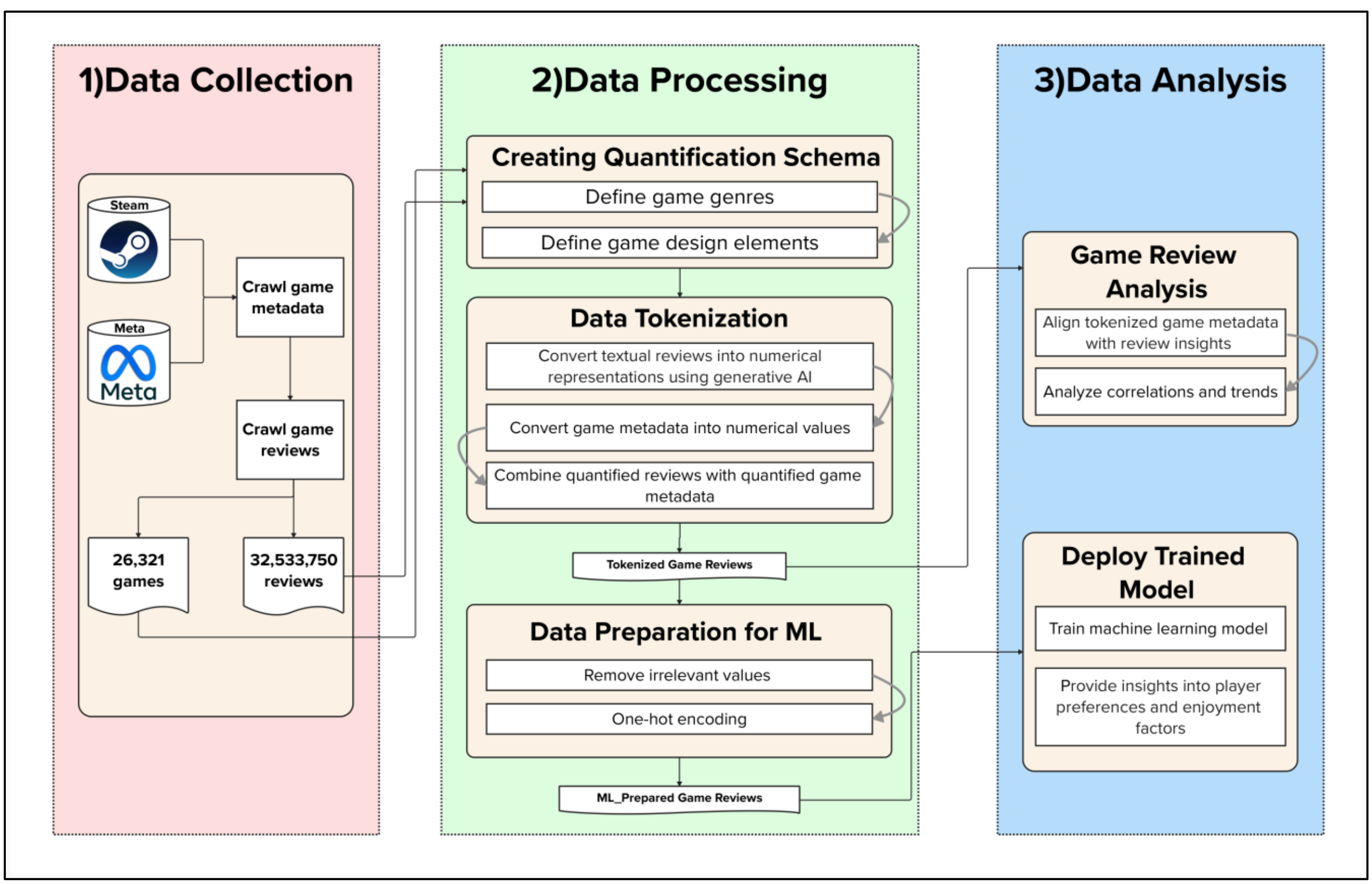

**Fig. 2** The workflow of the study

### 3.2 Data scraping

For Steam games, the process starts by loading previous datasets to avoid redundant requests. The script then retrieves a list of all games from Steam's API, using unique game IDs (AppIDs) as references. For each game, it checks if the game has already been processed or if it remains unreleased, in which case it is skipped. If a game is new and relevant, a request to the Steam API retrieves details shown in Fig. 3 like the game's name, release date, genres, and price.

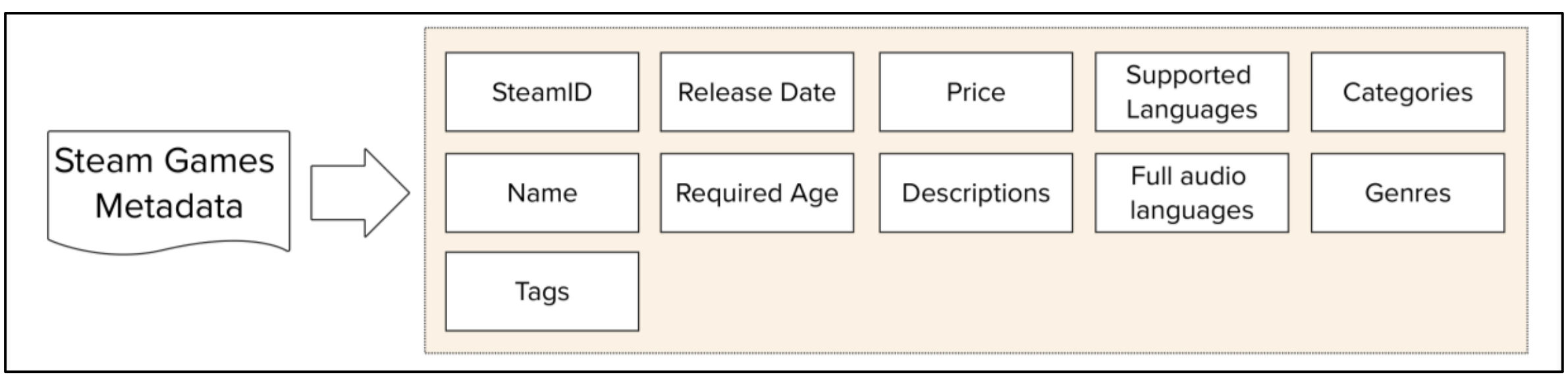

**Fig. 3** The source (raw) Steam games metadata dataset

Similarly, for VR games from the Meta Quest store, the process starts by accessing every game's unique URL from VRDB, a comprehensive database for Meta Quest VR games on the Meta Quest store [25]. The script makes HTTP requests to VRDB's website to retrieve pages of game listings, utilizing pagination to scrape multiple pages efficiently. Once fetched, the HTML content is parsed using BeautifulSoup to locate embedded JavaScript data containing game details. Using regular expressions, key attributes such as game ID, name, and genres are extracted, while additional details like release date, ratings, and price are captured, as can be seen in Fig. 4.

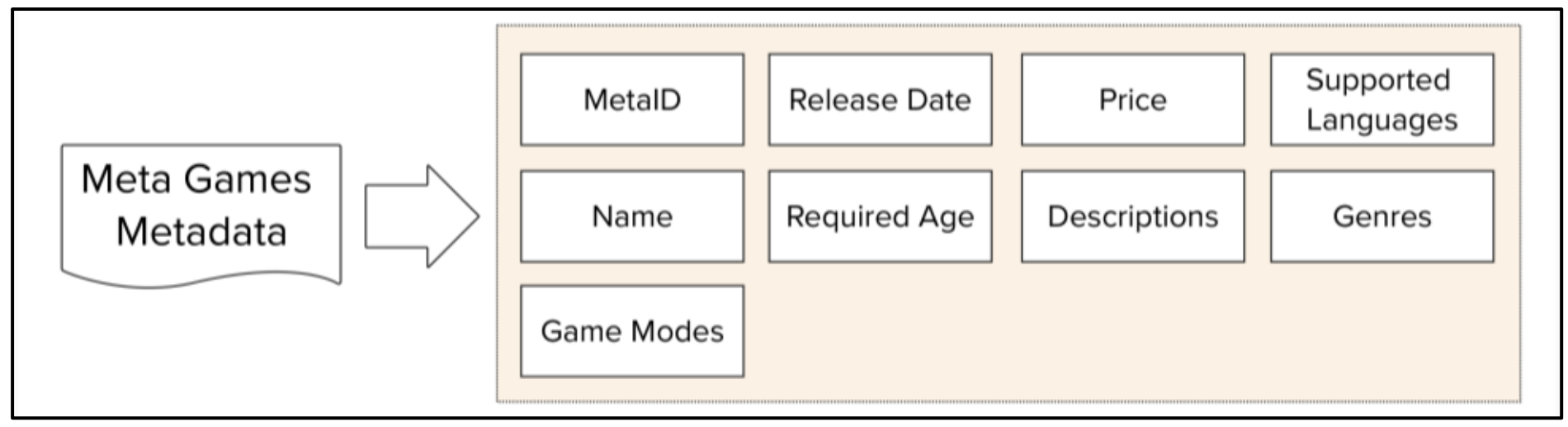

**Fig. 4** The source (raw) Meta games metadata dataset

To ensure accuracy, the script filters out games with release dates of 2019 or earlier, gathering only data for games that meet the specified date criteria. The extracted information is then saved to a JSON file, with regular autosaves to prevent data loss. Unreleased games or those with insufficient data are logged separately to streamline future runs.

By focusing on games with release dates starting from 2020, the scraper maintains a targeted dataset, storing only high-priority information and minimizing unnecessary API calls. The script's systematic approach ensures efficient, relevant data collection while maintaining data integrity and completeness.

Once all games are scraped, the review scraping script is executed to collect reviews for each game. The process begins by loading each game's ID and details from the raw games metadata file. It then verifies whether each game has a minimum of 25 reviews to process, ensuring that games with a low number of reviews are excluded to avoid potential biases. If the criterion is met, the script retrieves reviews in all available languages. This approach ensures the creation of a culturally diverse and generalized dataset, mitigating biases towards any specific culture or demographic. For each game with sufficient reviews, a CSV file is generated. These files are named according to the game ID and the total number of reviews scraped, enabling the incremental storage of all collected reviews. Reviews are written in order of rating, with the highest-rated review appearing in the first row. The Python process stops collecting reviews for a game once all reviews are gathered or the specified total review count is reached. This batch-oriented design facilitates efficient and organized data collection across multiple games and languages.

Tables 2 and 3 present the output of a custom Python script that processed individual CSV files to generate a descriptive summary. Of 65,686 Steam games, reviews were collected for 23,107 that met the minimum criterion of at least 25 reviews, totaling 31,832,390 reviews. Similarly, of 9,336 VR games on Meta, reviews were collected for 3,210, totaling 701,360 reviews. The resulting dataset of Steam games is openly accessible at Mendeley Data (https://doi.org/10.17632/jxy85cr3th.2) [26], while the dataset from the Meta Quest store is available upon request.

**Table 2** Descriptive summary of Steam store reviews dataset

| Store | Total number of games | Total sum of game reviews | Top 10 games with the highest review count |
|---|---|---|---|
| Steam | 23,107 | 31,832,390 | 1. Game ID: 553850 - 336,460 reviews<br>2. Game ID: 2357570 - 313,125 reviews<br>3. Game ID: 990080 - 293,388 reviews<br>4. Game ID: 1293830 - 263,270 reviews<br>5. Game ID: 261550 - 253,629 reviews<br>6. Game ID: 1468810 - 224,207 reviews<br>7. Game ID: 1599340 - 201,079 reviews<br>8. Game ID: 526870 - 194,630 reviews<br>9. Game ID: 1203220 - 186,800 reviews<br>10. Game ID: 427520 - 180,898 reviews |

**Table 3** Descriptive summary of Meta store reviews dataset

| Store | Total number of games | Total sum of game reviews | Top 10 games with the highest review count |
|---|---|---|---|
| Meta | 3,210 | 701,360 | 1. Game ID: 7190422614401072 - 9,731 reviews<br>2. Game ID: 4416817901687071 - 9,705 reviews<br>3. Game ID: 3947713001963656 - 9,692 reviews<br>4. Game ID: 7276525889052788 - 9,650 reviews<br>5. Game ID: 3661420607275144 - 9,646 reviews<br>6. Game ID: 2370815932930055 - 9,470 reviews<br>7. Game ID: 6061406827268889 - 9,465 reviews<br>8. Game ID: 4979055762136823 - 9,188 reviews |

9. Game ID: 4215734068529064 - 9,172 reviews
10. Game ID: 2031826350263349 - 9,115 reviews

### 3.3 Data processing

#### 3.3.1. Data quantification schema

The tokenization schema created for processing game and review data ensures systematic analysis by leveraging industry standards and insights from reputable sources. Age ranges are classified using the PEGI (Pan-European Game Information) rating system [27], while price ranges are based on VGinsights Steam Analytics [28], as shown in Fig. 5. These price ranges are divided into five categories:

- $0.01–$4.99 (below the ±$5 range for indie games).
- $5–$14.99 (within the ±$5 range for indie games).
- $15–$24.99 (within the ±$5 range for AA games).
- $25–$39.99 (within the ±$5 range for AAA games).
- $40+ (above the ±$5 range for Premium AAA games).

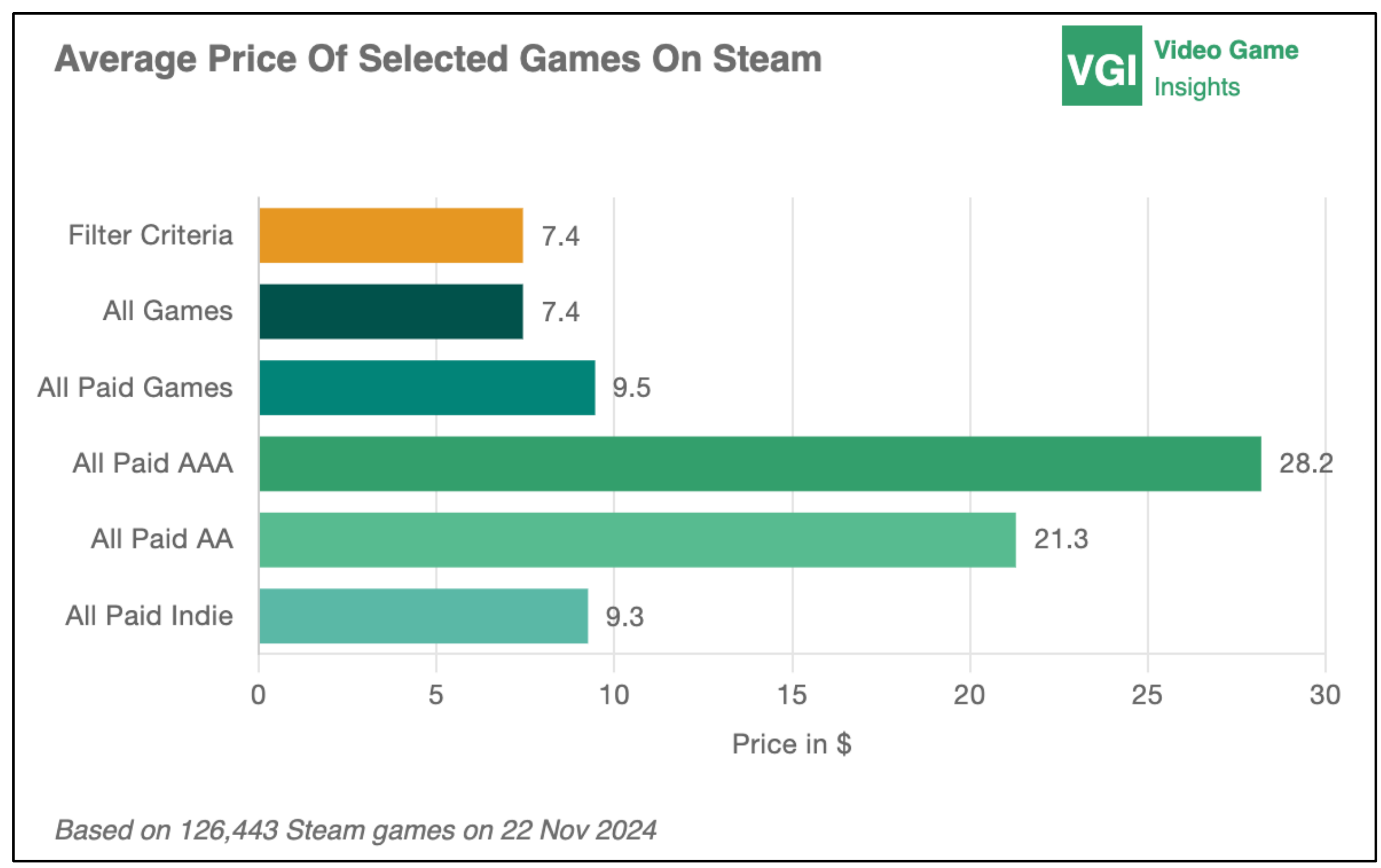

**Fig. 5** Average price of selected games on Steam [28]

Also, game genres are classified based on [4], [29], [30], and include Action, Adventure, Casual, Puzzle, Role-Playing Game (RPG), Racing, Simulation, Sports, Strategy, Fighting, Horror, Battle Royale, Shooter, Survival, Music, Education, Entertainment, Meditation, and Exercise. As shown in Table 4, these genres ensure comprehensive coverage of both PC and VR games. The 'Value' column in Table 4 represents the presence of a given genre in a game, where 0 indicates its absence and 1 signifies its inclusion. This classification allows for a structured analysis of game genres by the generative AI model.

**Table 4** Game genres tokenization schema

| Game Genre | Value | Description |
|---|---|---|
| Action | 0-1 | Fast-paced gameplay focused on combat and movement. |
| Adventure | 0-1 | Exploration and storytelling with player-driven narratives. |
| Casual | 0-1 | Simple, easy-to-learn gameplay often for short sessions. |
| Puzzle | 0-1 | Problem-solving challenges requiring logic and strategy. |
| Role-Playing | 0-1 | Character progression through quests and narrative-driven gameplay. |
| Racing | 0-1 | Competitive gameplay involving speed and vehicles. |
| Simulation | 0-1 | Realistic emulation of real-world activities or systems. |
| Sports | 0-1 | Competitive games based on physical sports. |
| Strategy | 0-1 | Planning and tactical decision-making gameplay. |
| Fighting | 0-1 | One-on-one or group combat with specific skill sets or techniques. |
| Horror | 0-1 | Games designed to evoke fear and suspense. |
| Battle Royale | 0-1 | Multiplayer survival games with a shrinking play area. |
| Shooter | 0-1 | Gameplay focused on firearms or ranged combat mechanics. |
| Survival | 0-1 | Resource management and survival in challenging environments. |
| Music | 0-1 | Games focused on rhythm and music interaction. |
| Education | 0-1 | Games designed to teach or reinforce learning objectives. |
| Entertainment | 0-1 | Games designed primarily for relaxation and fun without specific goals. |

Furthermore, the schema incorporates critical game design elements derived from [31], [32], [33], covering ‘Gameplay’, ‘Graphics’, ‘Difficulty’, ‘Story’, ‘Audio’, ‘Avatar Customization’, ‘Controls’, ‘Monetization Model’, ‘Replayability’, ‘Community’, ‘Multiplayer’, and ‘Spatial Presence’.

**Table 5** Game design elements tokenization schema

| Design Element | Value | Description |
|---|---|---|
| Gameplay | 1-5 | Score for the quality of core mechanics and interactive experience of the game. |
| Graphics | 1-5 | Score for the quality of visual presentation and design |
| Difficulty | 1-5 | Score for the quality of balance and challenge offered to players |
| Story | 1-5 | Score for the quality of narrative depth and engagement |
| Audio | 1-5 | Score for the quality of sound effects, music, and overall auditory design |
| Avatar Customization | 1-5 | Score for the quality of representation and customization of the player’s character |
| Controls | 1-5 | Score for the quality of usability and responsiveness of input methods |
| Monetization Model | 1-5 | Score for the quality of strategies used for game revenue, such as in-app purchases |
| Replayability | 1-5 | Score for the quality of replay value and potential for repeated play sessions |
| Community | 1-5 | Score for the quality of interaction and engagement with other players |
| Multiplayer | 1-5 | Score for the quality of features supporting cooperative or competitive play |
| Spatial Presence | 1-5 | Score for the quality of immersive feeling of being "inside" the game environment |

As shown in Table 5, these design elements are essential in assessing the overall quality of a game, as they influence player engagement, immersion, and enjoyment. Each game review is scored on a 1-5 scale against these design elements, allowing for a structured evaluation of gameplay mechanics, difficulty, graphics, audio, and spatial presence, which contribute to the overall experience. Storytelling enhances emotional investment, while customization options enable greater player expression. Additionally, monetization models, multiplayer features, and community engagement shape long-term retention and player perception. By evaluating these aspects, the schema provides a comprehensive framework for understanding how good or bad a game is.

Building on the framework established by Lin and his team [6], we further identified distinct categories to classify each review effectively, as shown in Table 6. These attributes allow for a structured analysis of review content, capturing aspects such as helpfulness, sentiment, suggestions, and technical issues. Additionally, we introduced the ‘Recommended’ and ‘Review Language’ attributes to enhance the classification schema. The ‘Recommended’ attribute provides insight into the overall sentiment of the reviewer, indicating whether they endorse the game. Meanwhile, ‘Review Language’ was added to account for linguistic diversity in user feedback, ensuring a more comprehensive understanding of reviews across different regions and player demographics.

The ‘Value’ column in Table 6 represents whether a specific review characteristic is present, with 0 indicating its absence and 1 signifying its inclusion. For ‘Review Language’, values range from 1 to 11, denoting different languages to facilitate multilingual analysis. By incorporating these elements, the schema enables a more nuanced evaluation of player reviews, supporting deeper insights into user experiences and preferences.

**Table 6** Review attributes tokenization schema

| Field Name | Value | Description |
|---|---|---|
| Recommended | 0-1 | Whether the reviewer recommends the game (1) or not (0) |
| Is_Helpful | 0-1 | Whether the content of the review is helpful to the developer (1) or not (0) |
| Is_Pro | 0-1 | Whether the review highlights a positive aspect of the game (1) or not (0) |
| Is_Con | 0-1 | Whether the review identifies a negative aspect of the game (1) or not (0) |
| Is_Video | 0-1 | Whether the review includes a URL to a video review (1) or not (0) |
| Is_Suggestion | 0-1 | Whether the review includes a suggestion on improving the game (1) or not (0) |
| Is_Bug | 0-1 | Whether the review describes a bug that occurs in the game (1) or not (0) |
| Review_Language | 1-11 | 1 = Simplified Chinese, 2 = English, 3 = Russian, 4 = Spanish, 5 = Portuguese, 6 = German, 7 = Japanese, 8 = French, 9 = Polish, 10 = Turkish, 11= Others |

The language categories were determined based on the top 10 most spoken languages on Steam, as identified in Steam's October 2024 Monthly Survey shown in Fig. 6. These include Simplified Chinese, English, Russian, Spanish, Portuguese, German, Japanese, French, Polish, and Turkish. Additionally, a separate category, 'Other' was included to account for reviews written in languages outside of this list, ensuring comprehensive language classification within the schema.

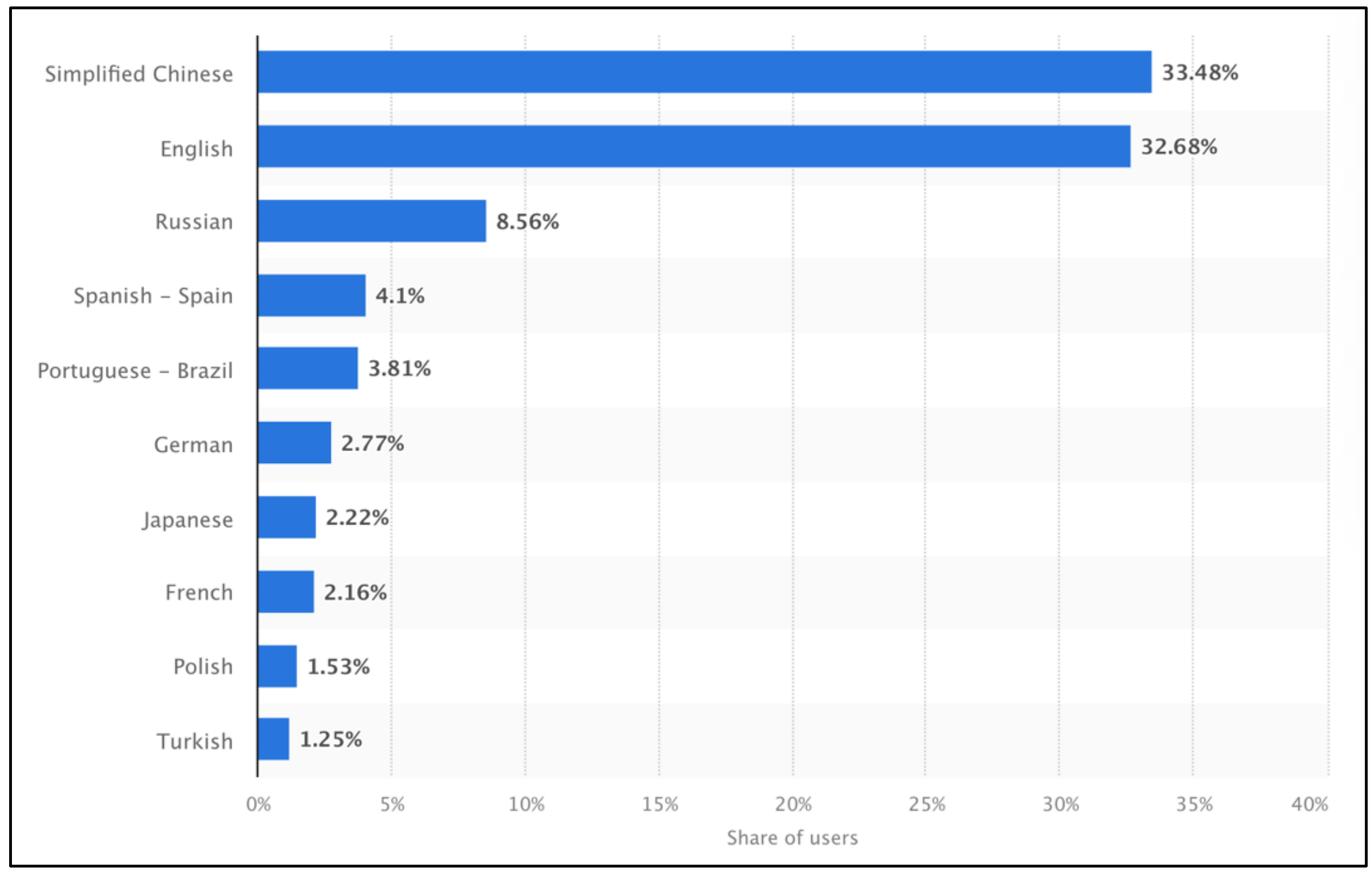

**Fig. 6** Steam's most common languages [34]

### 3.3.2. Quantifying game metadata

This process involves converting raw game metadata into structured categorical and numerical features. The goal is to create a uniform dataset where each game is represented by a set of predefined attributes. For both Meta Quest and Steam stores, the same schema is applied for feature extraction and tokenization. This ensures that game data from different platforms is processed in a unified manner, facilitating cross-platform analysis.

As shown in Tables 7 and 8, categorical mapping is applied to games using a set of predefined categories, each associated with relevant keywords. These keywords are extracted directly from the raw scraped games files, where they appear under the respective key names listed in the 'Derived From' column. For example, on the Steam store, the 'Action' category includes terms like 'Action RPG', 'Action Roguelike', 'Beat 'em up', and 'Hack and Slash', which are found in the 'Genre' or 'Tag Mapping' fields in the game's metadata [35].

**Table 7** Categorical mapping for Steam games

| Category | Keywords | Derived From |
|---|---|---|
| Action | Action, Action RPG, Action Roguelike, Action-Adventure, Beat 'em up, Character Action Game, Combat, FPS, Fighting, Hack and Slash, Shooter, Swordplay, Twin Stick Shooter | Genre, Tag Mapping |
| Adventure | Action-Adventure, Adventure, Atmospheric, Exploration, Fantasy, Open-world, Puzzle, Story Rich, Visual Novel | Genre, Tag Mapping |
| Battle_Royale | Battle Royale, FPS, Multiplayer, PvP, Shooter, Survival | Genre, Tag Mapping |
| Casual | 2D, Casual, Cute, Family Friendly, Free-to-play, Indie, Puzzle, Relaxing, Simulation | Genre, Tag Mapping |
| Education | Education, Game Development, Programming, Software Training, Trivia, Web Publishing | Genre, Tag Mapping |
| Entertainment | Entertainment, Feature Film, Movie, Streaming, TV, Video Production | Genre, Tag Mapping |
| Fighting | 2D Fighter, 3D Fighter, Beat 'em up, Boxing, Competitive, Fighting, Martial Arts, Wrestling | Genre, Tag Mapping |
| Free_To_Play | Free-to-play | Category Flag, Tag Mapping |
| Horror | Dark, Gore, Gothic, Horror, Lovecraftian, Psychological Horror, Survival Horror, Thriller, Vampires, Zombies | Genre, Tag Mapping |
| Is_3D | 3D | Tag Mapping |
| Is_Coop | Co-op, Online Co-op | Category Flag |
| Is_Early_Access | Early Access | Category Flag |
| Is_Indie | Indie | Category Flag |
| Is_Multiplayer | MMO, Multi-player, Multiplayer, Online PvP, PvP | Category Flag |
| Is_Singleplayer | Single-player | Category Flag |
| Is_Steam | Steam Achievements | Category Flag |
| Is_VR | VR Only, VR Support, VR Supported | Category Flag |
| Music | 8-bit Music, Electronic Music, Instrumental Music, Music, Music-Based Procedural Generation, Rhythm, Rock Music, Soundtrack | Genre, Tag Mapping |
| Puzzle | Card Game, Hidden Object, Logic, Match 3, Puzzle, Puzzle Platformer, Sokoban, Sudoku, Time Manipulation, Word Game | Genre, Tag Mapping |
| Racing | ATV, BMX, Combat Racing, Cycling, Driving, Motorbike, Offroad, Racing, Vehicular Combat | Genre, Tag Mapping |
| Role_Playing_Game | Action RPG, CRPG, Dungeon Crawler, JRPG, Party-Based RPG, RPG, Roguelike, Roguelite, Strategy RPG, Turn-Based RPG | Genre, Tag Mapping |
| Shooter | Arena Shooter, FPS, Gun Customization, Hero Shooter, Looter Shooter, On-Rails Shooter, Shooter, Tactical Shooter, Third-Person Shooter, Twin Stick Shooter | Genre, Tag Mapping |
| Simulation | City Builder, Colony Sim, Farming Sim, Immersive Sim, Management, Medical Sim, Simulation, Space Sim, Trading, Train Sim, Transportation | Genre, Tag Mapping |
| Sports | Baseball, Basketball, Boxing, Football (American), Football (Soccer), Golf, Hockey, Motocross, Pool, Racing, Rugby, Skateboarding, Sports, Tennis, Volleyball | Genre, Tag Mapping |
| Strategy | 4X, Card Battler, Chess, Deckbuilding, Grand Strategy, RTS, Real-time Tactics, Strategy, Strategy RPG, Tactical RPG, Tower Defense, Turn-Based Strategy, Turn-Based Tactics | Genre, Tag Mapping |
| Survival | Base Building, Crafting, Open-world Survival Craft, Resource Management, Roguelike, Roguelite, Survival, Survival Horror | Genre, Tag Mapping |

**Table 8** Categorical mapping for Meta games

| Category | Keywords | Derived From |
|---|---|---|
| Action | Action, Arcade, Fighting, Party Game, Platformer, Shooter | Genre |
| Adventure | Adventure, Narrative, Sandbox, World Creation | Genre |
| Battle_Royale | NA | NA |
| Casual | Hangout, Party Game, Platformer, Puzzle, Tabletop | Genre |
| Education | Learning | Genre |
| Entertainment | NA | NA |
| Fighting | Fighting | Genre |
| Free_To_Play | NA | Price |
| Horror | Survival | Genre |

| Is_3D | NA | NA |
|---|---|---|
| Is_Coop | Co-op | Game Mode |
| Is_Early_Access | NA | NA |
| Is_Indie | NA | NA |
| Is_Multiplayer | Multiplayer | Game Mode |
| Is_Singleplayer | Einzelspieler, Single User | Game Mode |
| Is_Steam | NA | NA |
| Is_VR | NA | NA |
| Music | Rhythm | Genre |
| Puzzle | Puzzle, Tabletop | Genre |
| Racing | Racing | Genre |
| Role_Playing_Game | Role Playing | Genre |
| Shooter | Shooter | Genre |
| Simulation | Sandbox, Simulation, World Creation | Genre |
| Sports | Sports | Genre |
| Strategy | Strategy, Tabletop | Genre |
| Survival | Sandbox, Survival, World Creation | Genre |

As noted in Table 8, some categories are not explicitly defined. This is because Meta does not provide specific labels for these categories. However, despite the absence of direct categorization, certain attributes can still be inferred from the raw game metadata.

For example, the 'Free_to_Play' category is determined based on the price value found in the raw game file. If the price is 0, the 'Free_to_Play' value is set to 1; otherwise, it is assigned 0. Similarly, since these games are sourced from the Meta Quest store, the 'Is_Steam' attribute is set to 0 by default, as they are not from Steam's store. On the other hand, the 'Is_VR' attribute is automatically set to 1, as all VR games are inherently 3D. The final output of this process is a CSV file in which each row represents a single quantified game.

### 3.3.3. Quantifying game reviews

To convert unstructured player reviews into analyzable features, we use Microsoft's Phi-4 small language model (SLM), a 14-billion-parameter Transformer-based model designed for high-quality reasoning with relatively modest computational requirements [36], [37]. Hosting Phi-4 locally allows us to process large volumes of review text while retaining control over the data. Figure 7 illustrates how Phi-4 achieves high-quality performance despite its relatively small size (14B parameters). Compared to both smaller and much larger models, Phi-4 demonstrates superior results on the Massive Multitask Language Understanding (MMLU) benchmark, highlighting its position at the frontier of "small but mighty" language models.

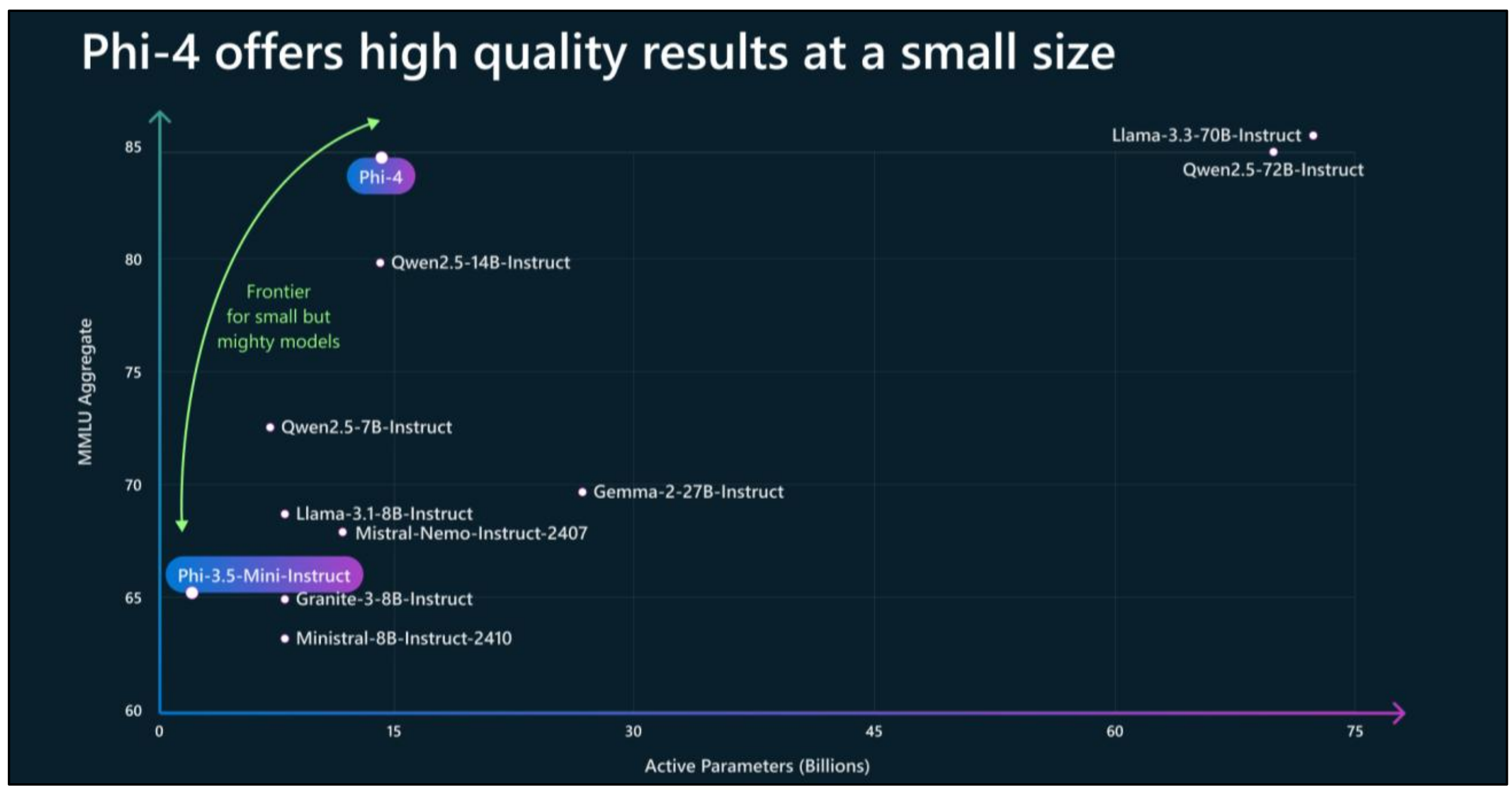

**Fig. 7** Phi-4 model benchmark [37]

Each review is mapped onto two complementary schemas. The first captures review-level attributes such as recommendation status, bug reports, suggestions, and language, as shown in Table 6. The second encodes ratings for 12 game-design elements including Gameplay, Graphics, Difficulty, and Spatial Presence, as defined in Table 5. The aim is to convert each free-text review shown in Fig. 8 into the standardized structure illustrated in Fig. 9, enabling all downstream analyses to treat reviews consistently as rows within a tabular dataset.

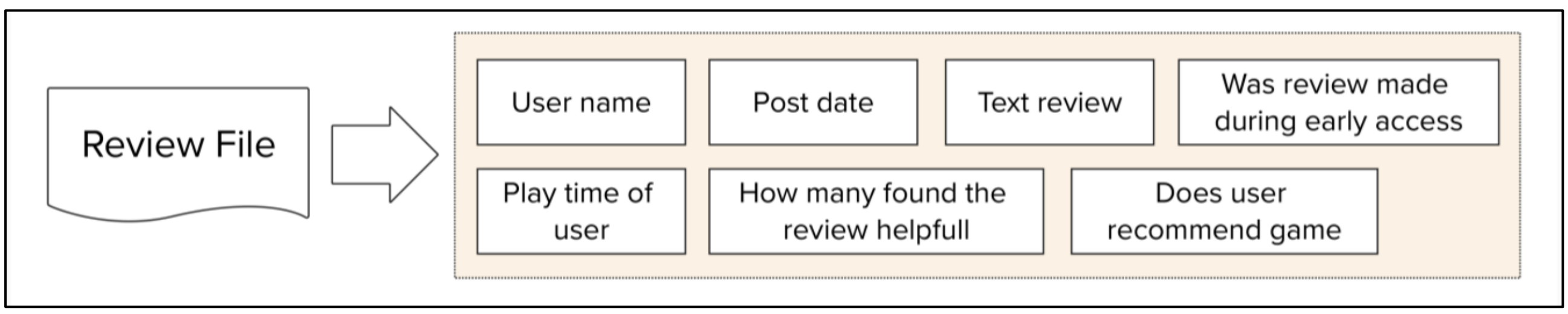

**Fig. 8** The source (raw) game review dataset

Only the reviews datasets were quantified, as the main file containing all games did not require generative AI quantification. Comparing the category columns in Figs. 8 and 9 reveals significant differences in rows, columns, and data values, adding complexity to the transformation process. The quantification approach employs generative AI for transformation, while column extraction was done manually based on literature. Although generative AI can suggest columns, adopting standard categories is generally preferable for comparability with other analyses.

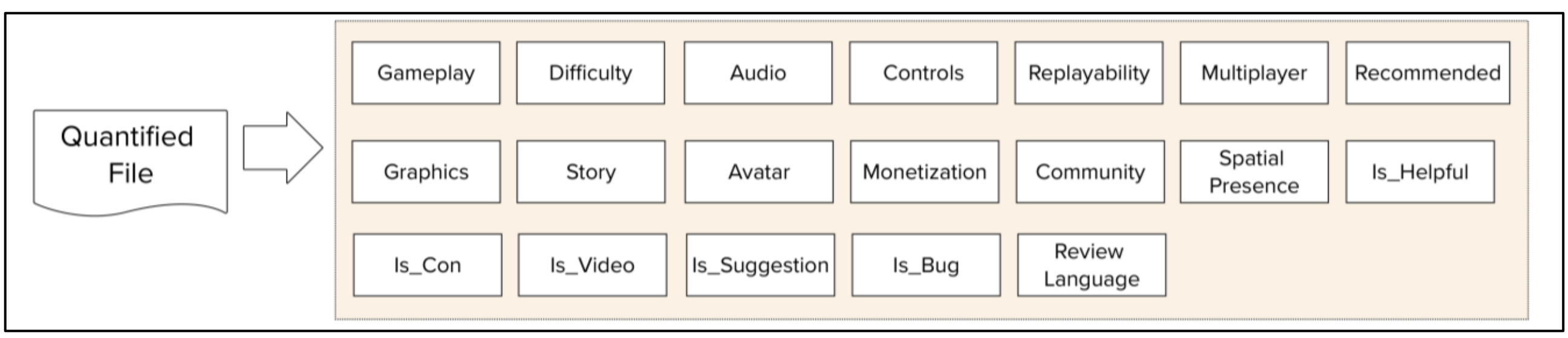

**Fig. 9** Structure of the target quantified review dataset

The comprehensive schemas in Tables 5 and 6 are processed using Microsoft's Phi-4 small language model hosted locally to tokenize data systematically, providing a robust foundation for advanced machine learning analysis.

The model quantifies the fields outlined in these schemas, assigning 0 to fields where there is insufficient data to evaluate. For instance, when the model processes a review such as "This game is amazing," it generates a structured JSON output as can be seen below in Table 9.

**Table 9** Structured JSON output from Phi-4 SLM

```
"Is_Helpful": 0,          "Recommended": 1,              "Controls": 0,
"Is_Pro": 1,              "Gameplay": 0,                 "Monetization_Model": 0,
"Is_Con": 0,              "Graphics": 0,                 "Replayablity": 0,
"Is_Video": 0,            "Difficulty": 0,               "Community": 0,
"Is_Suggestion": 0,       "Story": 0,                    "Multiplayer": 0,
"Is_Bug": 0,              "Audio": 0,                    "Spatial_Presence": 0
"Language": 2,            "Avatar_Customization": 0,
```

To achieve this, we've created a custom Python script. The script begins by defining both schemas highlighted in Tables 5 and 6 through a Pydantic model [38], which ensures consistent encoding of each text review's attributes and game design elements.

Table 10 shows how fields like 'Gameplay' and 'Graphics' are represented in a Pydantic model. This data processing model imposes strict type and range constraints, preventing incomplete or misleading values from entering the dataset.

**Table 10** Gameplay and Graphics schema in a Pydantic model

```
Gameplay: int = Field(
    ...,
    description=(
      "0–5 rating of gameplay quality:\n"
      "5 = (Excellent) Offers a wide range of engaging and well-integrated mechanics that ensure fluid interaction and adaptability.\n"
      "4 = (Very good) Mechanics are enjoyable and functional, with minor inconsistencies.\n"
      "3 = (Neutral) Solid mechanics, though limited in diversity or integration.\n"
      "2 = (Poor) Mechanics are clunky or lack coherence.\n"
      "1 = (Very poor) Gameplay is frustrating, with severe flaws.\n"
      "0 = (Unknown) Insufficient data to evaluate."
    )
  )
  Graphics: int = Field(
    ...,
    description=(
      "0–5 rating of graphics quality:\n"
      "5 = (Excellent) High fidelity, immersive environments with consistent visual quality.\n"
      "4 = (Very good) Impressive visuals with minor inconsistencies.\n"
      "3 = (Neutral) Decent visuals but lacks attention to detail or polish.\n"
      "2 = (Poor) Outdated or inconsistent graphics.\n"
      "1 = (Very poor) Distracting or poorly executed visuals.\n"
      "0 = (Unknown) Insufficient data to evaluate."
    )
  )
```

For each review, the script sends a prompt to the locally hosted Phi-4 SLM and expects a JSON object that conforms to the Pydantic schema. The prompt, summarized in Table 11, instructs the model to output integer-valued fields only. A LangChain-based wrapper converts the schema into format instructions embedded in the prompt so that the generated JSON can be parsed and validated automatically.

**Table 11** Prompt used to convert text reviews into numerical values in JSON format

```
Use the below schemas to convert any game text review into numerical values as per the schemas provided.
IMPORTANT: Your response MUST be exactly one valid JSON object, and nothing else.
Do not output multiple JSON blocks or repeated keys.
No comments, no trailing commas, no extra text.
Adhere to the following rules strictly:
1-If a review contains only symbols (e.g., "♥♥♥") or non-alphanumeric characters you should set all fields to 0 except for `Language`, which should be set to 11 (Other).
2-Do not infer positivity, negativity, or other attributes from special characters or emojis. Treat such reviews as uninformative unless meaningful words are present.
3-Ensure `Recommended` is strictly 0 or 1.
4-All rating fields (`Gameplay`, `Graphics`, etc.) must be integers between 0 and 5, in case you cant find sufficient data in the review that satisfies a specific field you should set that field to 0. Never give a 1 to 5 rating for any field if you cant find sufficient data in the review that supports it. NEVER MAKE UP DATA.
5-For binary fields (e.g., `Is_Helpful`, `Is_Pro`), ensure values are strictly 0 or 1.
6-If a game review includes numerical ratings (e.g., "gameplay is 7/10"), you should normalize these ratings to the required range of 1-5 where 0 represents insufficient data. Never use the ratings in the review directly without ensuring they abide to the schema. For example, a review that states "story is 7/10" would equate to setting the value for the 'Story' field to 4/5.

For example if text review is: "This game is amazing" your output would look like this:
"Is_Helpful": 0,"Is_Pro": 1,"Is_Con": 0,"Is_Video": 0,"Is_Suggestion": 0,"Is_Bug": 0,"Language": 2,"Recommended": 1,"Gameplay": 0,"Graphics": 0,"Difficulty": 0,"Story": 0,"Audio": 0,"Avatar_Customization": 0,"Controls": 0,"Monetization_Model": 0,"Replayablity": 0,"Community": 0,"Multiplayer": 0,"Spatial_Presence": 0

{format_instructions}

Actual review to be tokenized is: "{review}"
```

The SLM assigns scores to fields such as ‘Gameplay’, ‘Graphics’, and ‘Difficulty’, and classifies reviews by attributes like language and recommendation status. The script sanitizes text, and stores successfully parsed reviews with their original IDs and in a new CSV file. This workflow yields a transparent, machine-readable dataset that preserves data integrity while capturing qualitative insights from player reviews.

We quantified the 100 most upvoted reviews from a total of 4,856 PC and VR games. Specifically, we analyzed the top 3,860 games with the highest number of ratings from the Steam store and 996 from the Meta Quest store. By focusing on the most-rated reviews from games with the highest number of ratings, we aimed to ensure a representative analysis of player sentiment while minimizing biases that may arise from less-reviewed or niche titles.

### 3.3.4. Evaluation

In order to evaluate the proposed approach, we analyzed the top-rated 1,000 text reviews from the game HELLDIVERS 2, which had the highest number of reviews among all the games scraped. The tokenization process highlighted earlier in the report transformed these reviews into structured numerical data. This tokenized dataset, stored in a .csv file, was uploaded to Google Looker Studio [39] to generate visualizations.

Figure 10 illustrates the proportion of recommended versus non-recommended reviews, as well as the alignment between the SLM-based tokenization assessments and the actual raw data. The discrepancy between the SLM-based assessment and the actual recommendation data from Steam can be attributed to the fact that SLM-based assessment determines whether a review is positive or negative based on text sentiment analysis, which involves evaluating word choice, tone, and context. However, on Steam, users explicitly click ‘Recommended’ or ‘Not Recommended’, which may not always align with the sentiment expressed in the text.

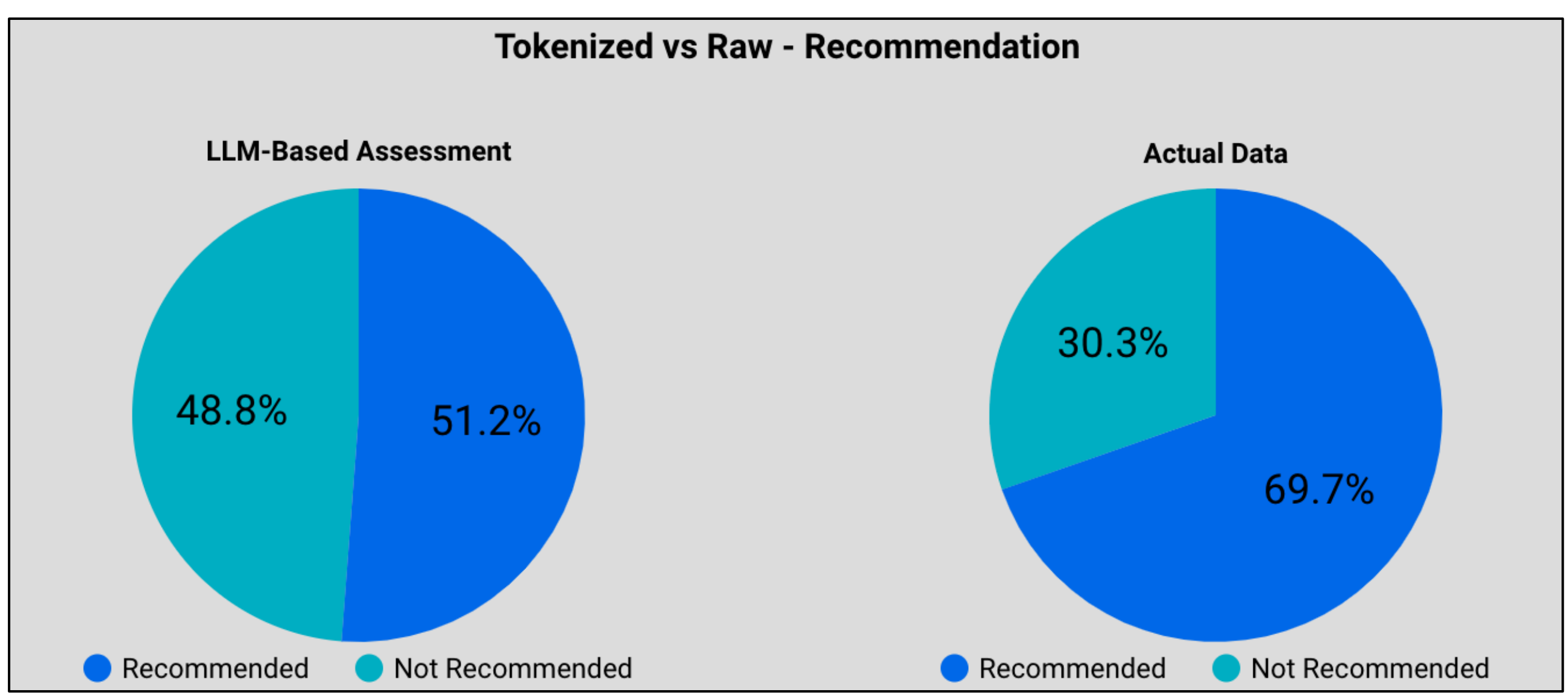

**Fig. 10** Recommendation column in tokenized vs raw files for HELLDIVERS 2

Figure 11 illustrates the Phi-4 model's attempt to label the languages of the top 1,000 highest-rated reviews. The results align with Gamalytic's statistics [40], which indicate that 40.2% of the player base originates from English-speaking countries, such as the United States and the United Kingdom, while 8.4% are from China.

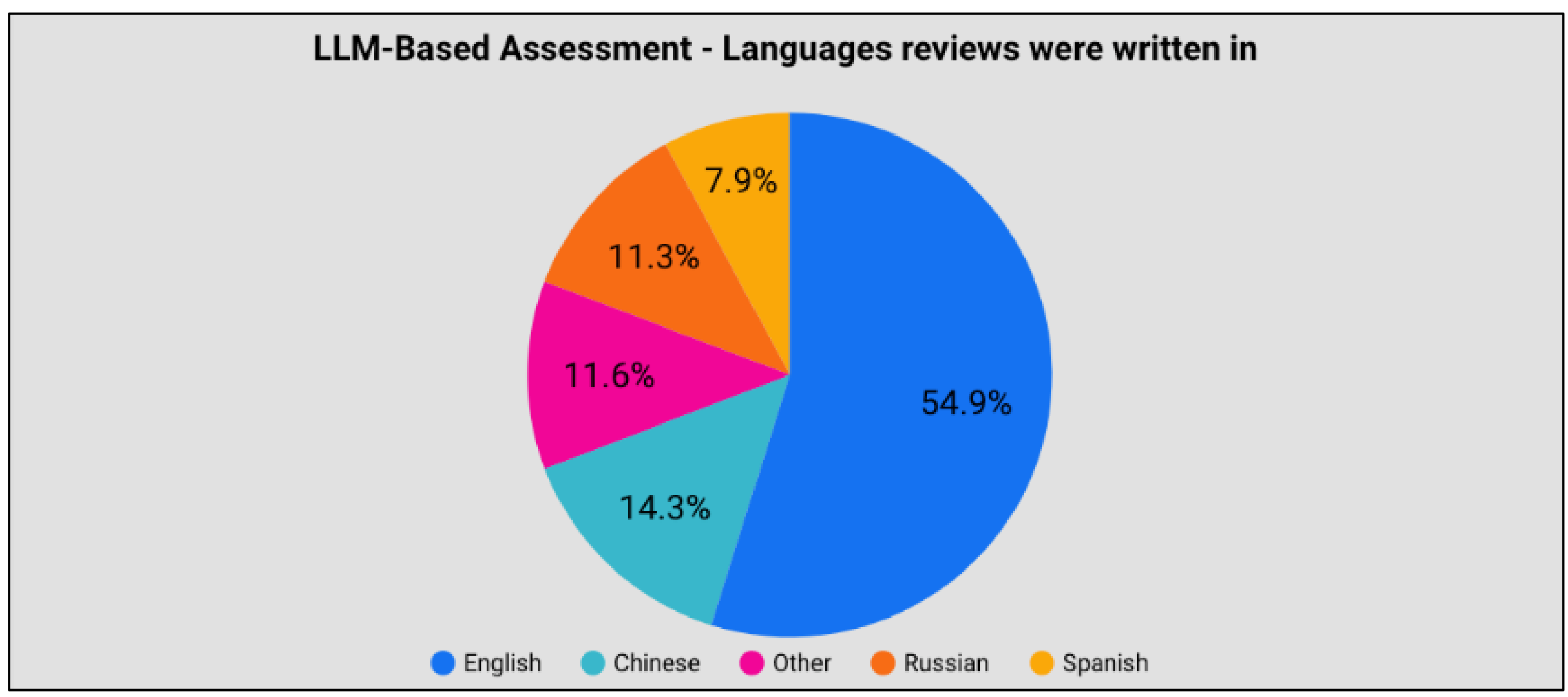

**Fig. 11** Language distribution of HELLDIVERS 2 text reviews

To evaluate the effectiveness of Phi-4 in quantifying qualitative player feedback, we applied it to the dataset of 1,000 textual reviews from HELLDIVERS 2. As seen in the bar charts in Fig. 12, Phi-4 successfully scored and quantified the reviews based on the 12 key game design elements shown in Table 5, demonstrating its ability to extract meaningful insights from unstructured text. The structured rating distributions suggest that the model is highly effective in translating subjective player experiences into a numerical assessment.

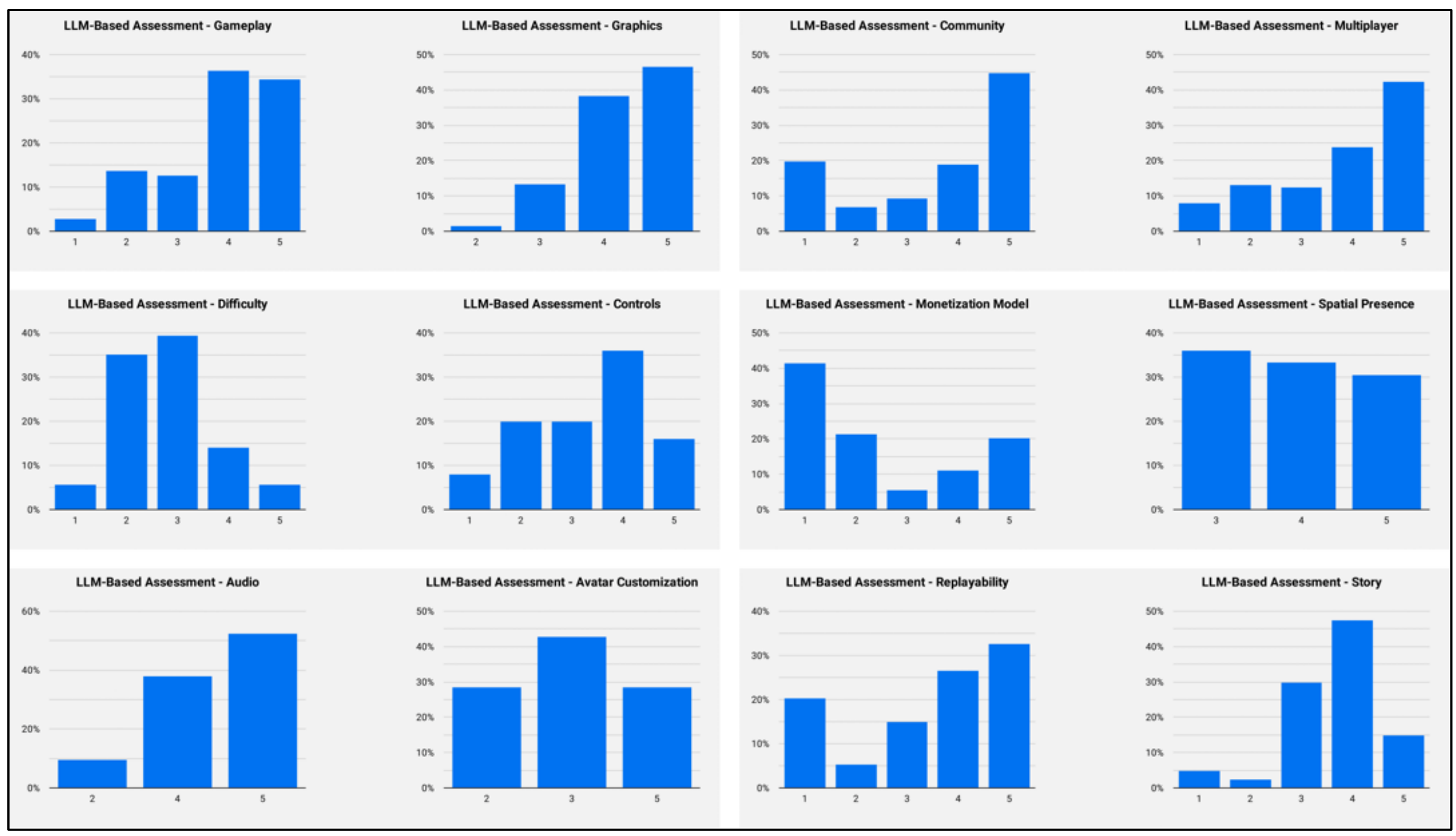

**Fig. 12** (Phi-4)-based rating distributions for the 12 game design elements in HELLDIVERS 2

For the 'Difficulty' element in HELLDIVERS 2, the bar chart on the left in Fig. 13 reveals that most ratings are concentrated around scores 2 and 3. This indicates that players generally perceive the difficulty as mostly fixed and unfair, as outlined in the schema (Table 5).

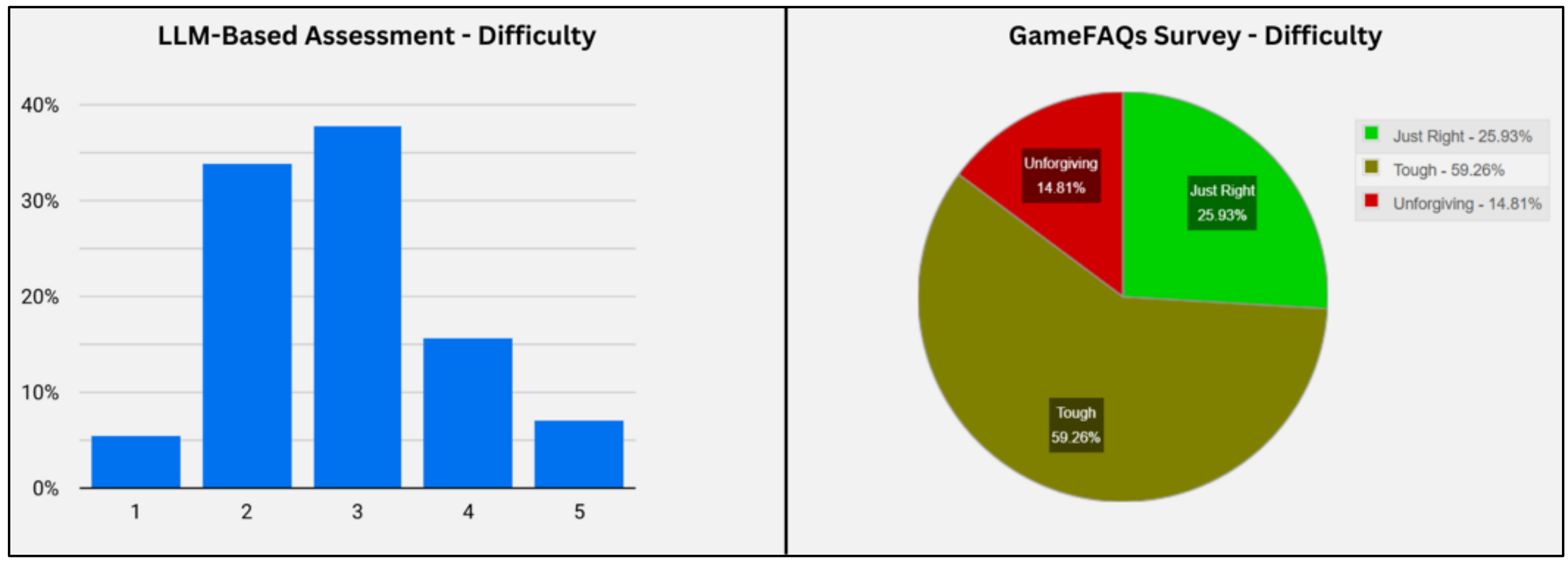

**Fig. 13** Distribution of 'Difficulty' ratings in (Phi-4)-based assessment vs. GameFAQs Survey

Interestingly, this assessment corresponds closely with data from the GameFAQs survey [41] which surveyed 46 players in total when it comes to the difficulty of HELLDIVERS 2 and the result can be seen in the pie chart on the right in Fig. 13. Both the SLM's assessment and the statistic from GameFAQs underscore the game's challenging nature. The GameFAQs data highlights subjective player sentiments such as 'Tough' or 'Unforgiving,' whereas Phi-4 focuses on structured evaluations of the difficulty design, including fairness and pacing. These perspectives are complementary, converging on the notion that the game is demanding and overly difficult. This underscores the validity of the model's evaluation, highlighting its alignment with player-reported perceptions of the game's difficulty.

From Fig. 14, it's evident that core elements such as 'Gameplay,' 'Graphics,' and 'Story' earn high ratings, while mid-range aspects like 'Difficulty' and 'Replayability' suggest room for improvement. 'Monetization ranks lowest, indicating that players enjoy the current approach least and view it as an area in need of enhancement. To provide an overall assessment of HELLDIVERS 2, an aggregate score of 3.5/5 was calculated by averaging the individual design element ratings shown in Fig. 14, aligning well with published ratings. Notably, this SLM-derived score does not conflict with the 6.6/10 overall user rating on Metacritic [42], highlighting a general consistency with user perceptions.

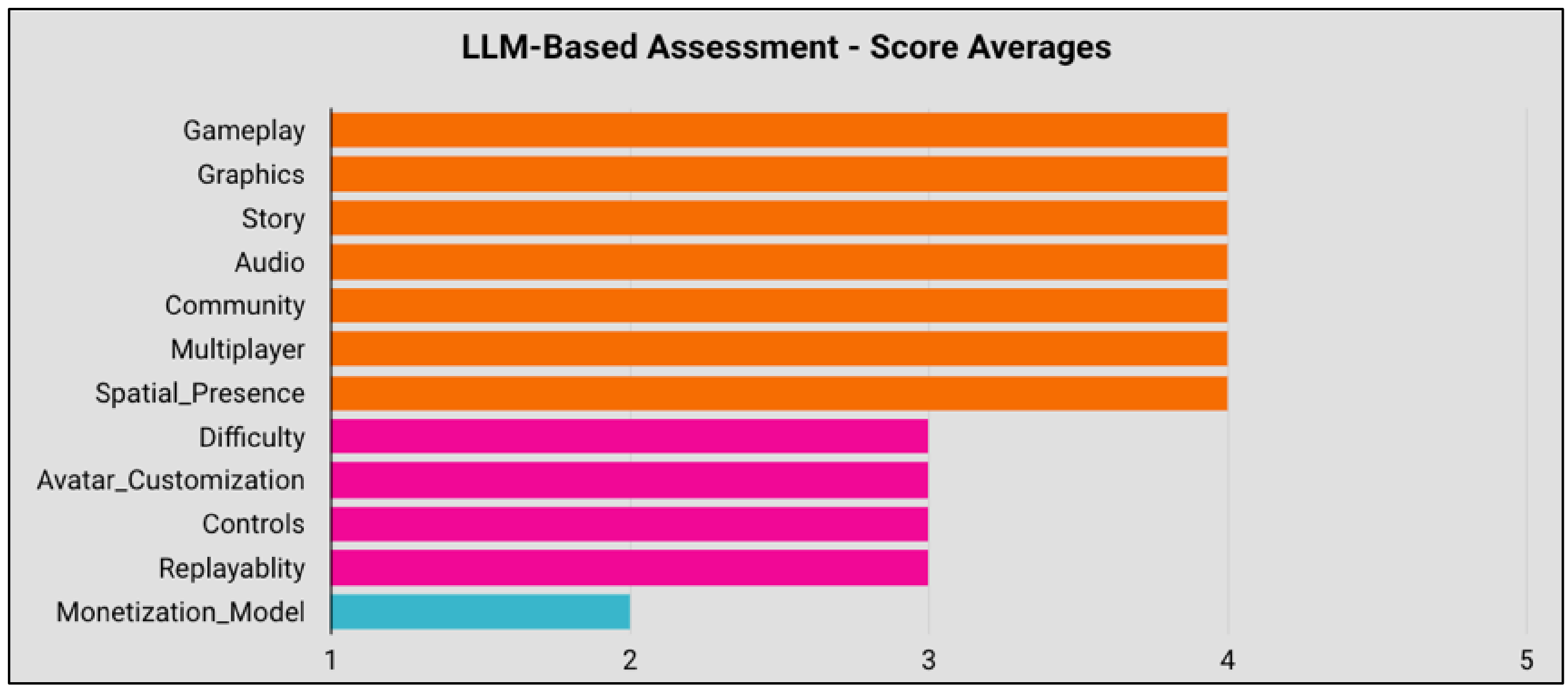

**Fig. 14** Averages of tokenized scores across game design elements for HELLDIVERS 2

### 3.4 Data preparation

Before conducting the analyses, we first prepared our quantified data. We began by combining the 4,856 individual quantified review files, each representing reviews for a single game, into a single dataset. The objective was to calculate the average rating for each game design element and store it as a single row per game in a CSV file named 'Tokenized_Reviews_Averages.csv'. Each of the 12 game design elements was rated on a 1-5 scale (1 = very low, 5 = very high). The elements are:

- Gameplay
- Difficulty
- Graphics
- Story
- Audio
- Avatar Customization
- Controls
- Monetization Model
- Replayability
- Community
- Multiplayer
- Spatial Presence

The Python script iterated through all review CSVs, calculated average scores per game, and merged them into a single dataset. This dataset was later uploaded to BigQuery for further analysis.

Finally, all game data was structured and stored in Google Cloud Storage, a cloud-based service for storing data [43]. This step was necessary to ensure that the dataset could be processed seamlessly, allowing for structured exploration and comparative analysis across different game platforms. The data uploaded to Google Cloud Storage consisted of two key datasets: 'Tokenized Game Metadata.csv' and 'Tokenized Reviews Data.csv', both of which served as the foundation for all subsequent analytical steps.

The next step was to use Google BigQuery, a platform that allows large-scale data analysis and management [21], to write queries on our data and derive insights and visualizations. The first action taken was to merge both datasets into a single table in BigQuery, with the price attribute transformed from a numerical value into a categorical attribute, as outlined in the Data quantification schema section (3.3.1). This transformation made it easier to run queries and analyze game metadata alongside review data.

### 3.5 Creating visuals in Google Looker

#### 3.5.1. Correlation between positive ratings and game design elements

Once we had our dataset in Google BigQuery, we began running queries to visualize the data differently. We created a query that calculates the correlation between price category and high rating percentages across different game design elements for both PC and VR games using the CORR() function in BigQuery.

First, the query converts Price_Category into numerical values (0 to 5), where 0 represents free-to-play games and 5 represents premium AAA games, enabling statistical correlation analysis. It then computes the percentage of high ratings (4+) across 12 game design elements (such as 'Graphics', 'Audio', 'Gameplay', and 'Replayability') and averages them to obtain a total high rating percentage for each game.

Using the CORR() function, the query determines the Pearson correlation coefficient between the encoded price category and the total high rating percentage, separately for VR games (Is_VR = 1) and PC (non-VR) games (Is_VR = 0). This analysis provides insights into whether pricing influences overall game ratings differently in VR and PC gaming environments.

#### 3.5.2. Comparing player enjoyment across VR and PC games

A BigQuery query was written to calculate the percentage of PC and VR games that receive high ratings (4+ on a 1-5 scale) across 12 game design elements. It first groups the dataset by Is_VR (where 1 represents VR games and 0 represents PC games) to analyze the difference in rating distributions between these two categories. For each design element, it computes the percentage of games with high ratings by dividing the count of games rated 4 or higher by the total number of games in that category. Additionally, the query calculates a total high rating percentage by averaging the high ratings across all 12 design elements, providing an overall metric for game quality in VR vs. PC games. This helps in identifying whether VR games tend to receive higher or lower ratings compared to non-VR games across different aspects of game design.

#### 3.5.3. Highest-rated game design elements by game genre

A BigQuery query was developed to analyze how different game genres influence player ratings across game design elements. The games metadata dataset originally stored genres as separate binary columns (Table 4), with a value of 1 indicating membership. To enable more flexible analysis, the query first unpivots these columns into a long-format structure, where each row represents a game–genre combination. This allows each game to be analyzed under multiple genres while preserving key metadata such as VR support, 3D capability, and multiplayer mode. The second stage unpivots game design elements, producing a dataset where each row corresponds to a rating for a game within a given genre. This structure supports detailed comparisons of how genres affect design aspect ratings, revealing whether some genres excel in specific areas. The final output provides a structured view of the interactions among game genre, platform type (VR vs. PC), and gameplay features.

To measure variances, we calculated standard deviations for both genre and platform (PC vs. VR). For each game design element, we first found the average rating for each genre (e.g., 'Gameplay' rating for all 'Action' games). Then, we used the STDDEV_SAMP function on these averages to see how much the ratings varied across genres. We did the same for platforms where ratings were grouped by platform, genre, and design element. Then we calculated platform-level averages and the standard deviation of the genre-level averages to understand how much player opinions varied within each platform. The standard deviation between platforms for a given design element is calculated as shown in (1). Here, $x_{PC}$ and $x_{VR}$ represent the mean ratings of the design element on PC and VR platforms, respectively. Since there are only two platforms $n = 2$, the general sample standard deviation formula reduces to the absolute difference between the two platform means divided by 2.

$$s = \frac{|x_{\mathrm{PC}} - x_{\mathrm{VR}}|}{\sqrt{2}} \tag{1}$$

### 3.5.4. Applying machine learning using XGBoost

To extend the analysis beyond descriptive statistics and explore predictive modeling, a machine learning approach was implemented using XGBoost, a high-performance gradient boosting algorithm [44]. XGBoost was selected for this study due to its ability to handle non-linear interactions, categorical variables, and missing data more effectively than traditional regression models [45]. A boosted tree model predicts the output $\hat{y}_i$ for an input $x_i$ as the sum of multiple decision trees as expressed in (2) where M is the total number of trees, $F_m(x_i)$ is the prediction from the m-th decision tree [45].

$$\hat{y}_i = F_M(x_i) = \sum_{m=1}^{M} f_m(x_i) \quad (2)$$

Game ratings are influenced by intricate dependencies between features, such as the interplay between price, genre, and multiplayer functionality. For instance, while VR games might receive high ratings for spatial presence, they do not necessarily score higher on community engagement. Similarly, higher-priced games may exhibit strong graphics quality but lower multiplayer engagement. These types of non-linear relationships are naturally handled by XGBoost's decision tree-based structure.

Additionally, XGBoost provides robustness against outliers [45], a common challenge in game rating analysis. Some free-to-play games, for example, receive extreme ratings, either overwhelmingly positive or negative, based on factors unrelated to gameplay quality, such as microtransactions. Linear Regression is highly sensitive to such extreme values, whereas XGBoost assigns less influence to outliers, improving prediction stability.

Lastly, XGBoost scales efficiently to large datasets, making it particularly well-suited for analyzing thousands of games and millions of player reviews. The dataset used in this study contained a diverse set of features with complex interdependencies, making XGBoost's ability to handle correlated features a critical advantage.

The objective is to predict overall game ratings based on game metadata, player reviews, and pricing categories. By leveraging XGBoost, we aim to uncover complex relationships between game attributes and player enjoyment, allowing for data-driven insights into what factors contribute to highly rated games. To ensure accurate predictions, we performed one-hot encoding for the Price Category attribute and created a new table without irrelevant attributes such as ID, Is Steam, Required Age, and Is Early Access. Table 12 presents the BigQuery query used to prepare the dataset for machine learning.

**Table 12** BigQuery query used to prepare dataset for machine learning

```
CREATE OR REPLACE VIEW `game_reviews.game_ml_dataset` AS
SELECT
  -- One-hot encode `Price_Category`
  CASE WHEN Price_Category = 'Free' THEN 1 ELSE 0 END AS Price_Free,
  CASE WHEN Price_Category = 'Low-Priced Indie' THEN 1 ELSE 0 END AS Price_Low_Priced_Indie,
  CASE WHEN Price_Category = 'Mid-Priced Indie' THEN 1 ELSE 0 END AS Price_Mid_Priced_Indie,
  CASE WHEN Price_Category = 'AA Games' THEN 1 ELSE 0 END AS Price_AA_Games,
  CASE WHEN Price_Category = 'AAA Games' THEN 1 ELSE 0 END AS Price_AAA_Games,
  CASE WHEN Price_Category = 'Premium AAA Games' THEN 1 ELSE 0 END AS Price_Premium_AAA_Games,
  -- User inputed features
  Is_VR, Is_3D, Is_Indie, Free_To_Play, Is_Coop, Is_Singleplayer, Is_Multiplayer,
  Action, Adventure, Casual, Puzzle, Role_Playing_Game, Racing,
  Simulation, Sports, Strategy, Fighting, Horror, Battle_Royale,
  Shooter, Survival, Music, Education, Entertainment
FROM `game_reviews.prediction_inputs`
```

Following this, we initiated the training of XGBoost by executing the query shown in Table 13 on the newly prepared dataset.

**Table 13** BigQuery query to run XGBoost regression

```
CREATE OR REPLACE MODEL `game_reviews.game_rating_predictor`
OPTIONS(model_type='BOOSTED_TREE_REGRESSOR',
    input_label_cols=['avg_overall_rating']) AS
SELECT
  -- One-hot encoded `Price_Category`
  Price_Free, Price_Low_Priced_Indie, Price_Mid_Priced_Indie,
  Price_AA_Games, Price_AAA_Games, Price_Premium_AAA_Games,
  -- Other categorical and binary features
  Is_VR, Is_3D, Is_Indie, Free_To_Play, Is_Coop, Is_Singleplayer,Is_Multiplayer,
  -- One-hot encode genres
  Action, Adventure, Casual, Puzzle, Role_Playing_Game, Racing,
  Simulation, Sports, Strategy, Fighting, Horror, Battle_Royale,
  Shooter, Survival, Music, Education, Entertainment,
  avg_overall_rating
FROM `game_reviews.game_ml_dataset`;
```

## 4 Analysis

### 4.1 Overview of quantified data

Figure 15 presents the top 10 genre distribution for PC vs. VR games, showcasing the percentage of the quantified dataset of games metadata occupied by each genre across the two platforms. The left side of the figure illustrates PC games, while the right side highlights VR games. The distributions are based on the total dataset of 4,856 games, capturing the most commonly occurring genres in each platform.

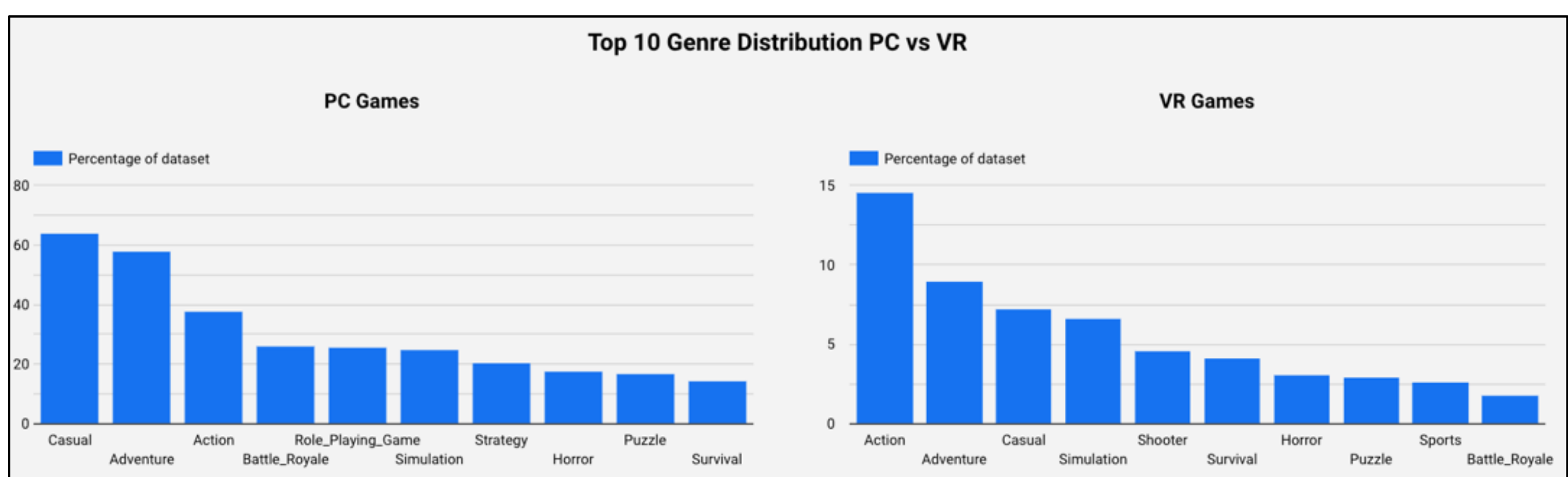

**Fig. 15** Genre distribution of PC vs VR Games

In PC games, the most prevalent genre is Casual, which makes up the largest portion of the dataset, followed closely by Adventure. Action games hold the third-highest share, while genres such as Battle Royale, Role-Playing Games, and Simulation have a moderate presence. The remaining genres—Strategy, Horror, Puzzle, and Survival—appear less frequently but are still among the most common.

In VR games, Action emerges as the most dominant genre, significantly surpassing other categories. Adventure and Casual games also hold substantial shares but trail behind Action. Simulation follows closely, while Shooter, Survival, and Horror genres show moderate representation. The lowest-ranking genres in this top 10 list include Puzzle, Sports, and Battle Royale.

This distribution provides an essential context for the data analysis, offering insight into how genre representation differs between PC and VR platforms. The differences in distribution suggest that certain genres are more prevalent in VR than in PC and vice versa.

### 4.2 Correlation between price and players' positive ratings

#### 4.2.1. PC games

Figure 16 presents the correlation values between various game design elements and price in PC games. 'Avatar Customization' (0.97) and 'Audio' (0.96) exhibit the highest positive correlations with price, meaning that as game prices increase, these features are rated more favorably. 'Spatial Presence' (0.94) and 'Replayability' (0.85) also show strong positive correlations, followed by 'Graphics' (0.62), 'Controls' (0.37), and 'Gameplay' (0.36), which maintain moderate positive relationships with price.

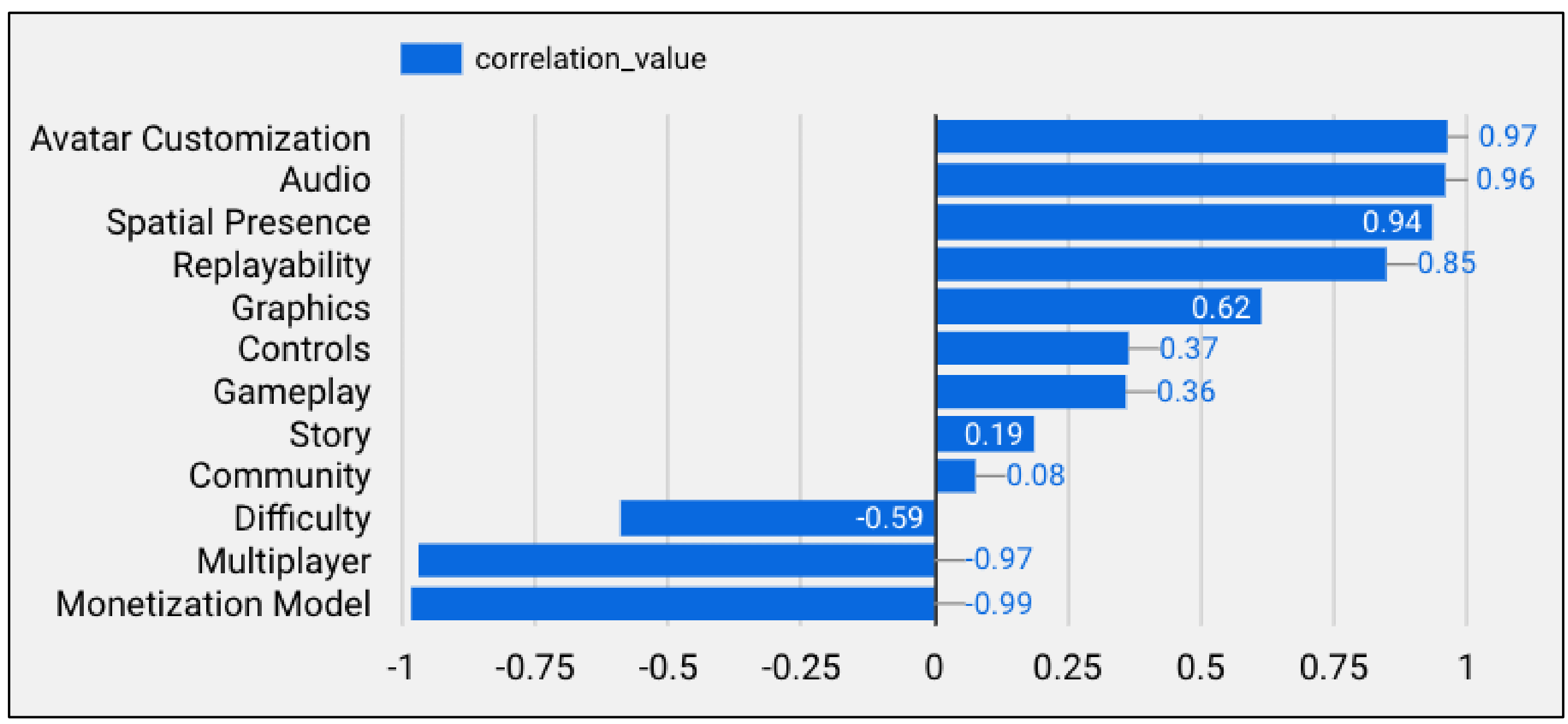

**Fig. 16** Correlation between positive ratings and game design elements in PC games

Story (0.19) and Community (0.08) display weak positive correlations, indicating that these aspects have little direct association with price. Difficulty (-0.59) is the first feature with a negative correlation, meaning that as game price increases, difficulty ratings tend to decrease. Multiplayer (-0.97) and Monetization Model (-0.99) have the strongest negative correlations, suggesting that these elements are rated lower in higher-priced games.

#### 4.2.2. VR games

As shown in Fig. 17, in VR games, 'Spatial Presence' (0.93) exhibits the highest positive correlation with price, meaning that more expensive VR games tend to provide significantly better immersive experiences. 'Story' (0.86) and 'Audio' (0.84) also show strong positive correlations with price, suggesting that higher-budget VR games are more likely to feature engaging narratives and high-quality sound design. 'Graphics' (0.82), 'Controls' (0.79), and 'Replayability' (0.64) follow closely behind, maintaining notable positive relationships with price. 'Gameplay' (0.55) and 'Multiplayer' (0.54) also correlate positively with price, though to a slightly lesser degree.

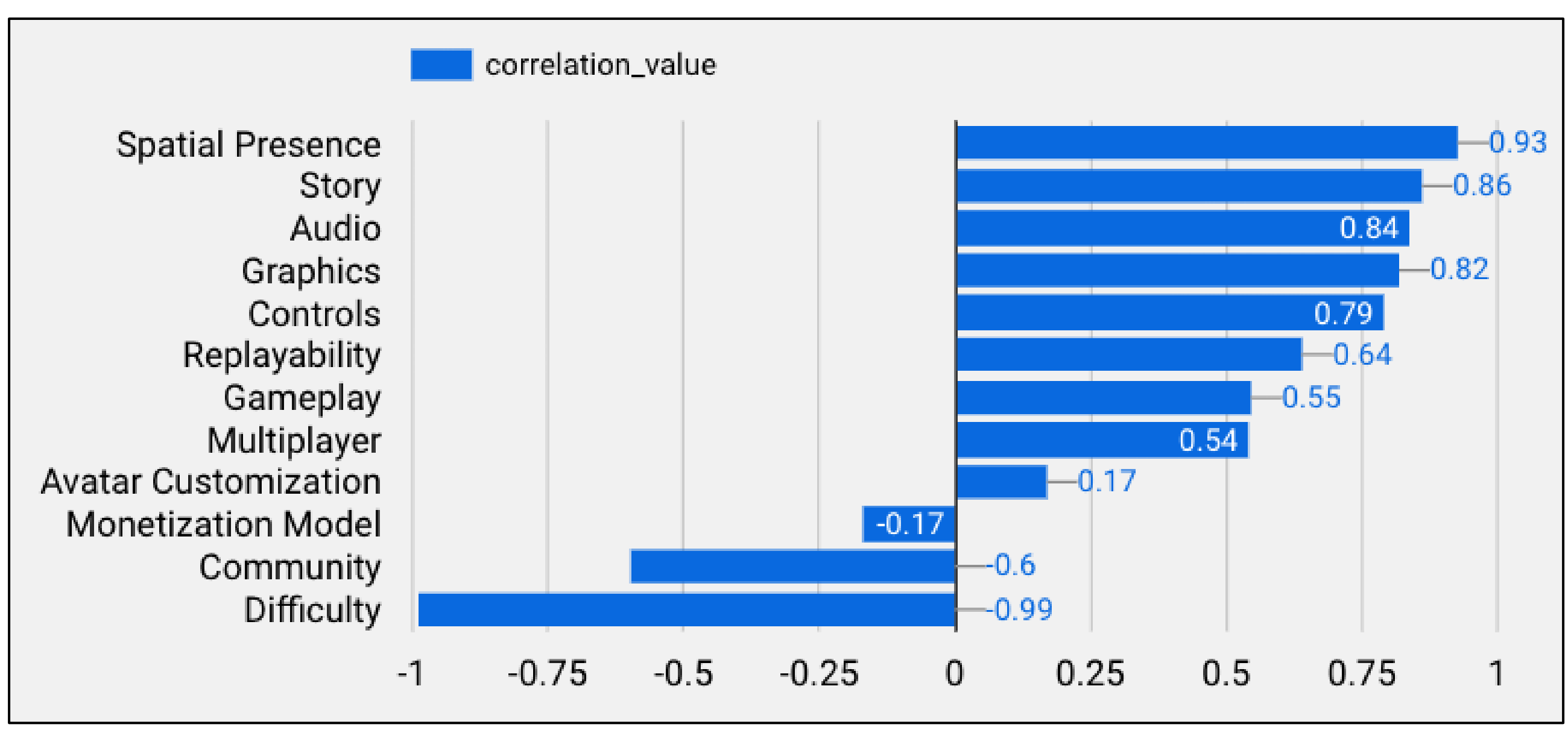

**Fig. 17** Correlation between positive ratings and game design elements in VR games

Monetization Model (-0.17) and Community Engagement (0.06) exhibit minimal correlation with price, indicating that monetization strategies and community-driven aspects do not strongly depend on a game's cost in VR. Difficulty (-0.99) has the strongest negative correlation with price, meaning that as the price of a VR game increases, difficulty ratings tend to decrease significantly.

#### 4.2.3. PC vs VR games

The comparison of Fig. 16 and Fig. 17 highlights key differences in how price correlates with player ratings across various game design elements in PC and VR games. While both platforms share some similarities, notable differences emerge in how game price influences design priorities and player reception.

One of the most significant differences is in the correlation between price and Story, which is high in VR games (0.86) but low in PC games (0.19). This means that higher-priced VR games are strongly associated with better storytelling, whereas story quality in PC games does not show a meaningful relationship with price. Audio quality, on the other hand, shows a strong correlation with price in both PC (0.96) and VR (0.84) games, indicating that sound design is consistently valued in premium titles regardless of platform.

Calculating the average percentage of high ratings across 12 game design elements, and computing the correlation between price and ratings separately for VR and PC games gave us two correlation values that can be seen in Fig. 18. We can see that VR games have a much higher total positive correlation (0.8) with price compared to PC (0.5).

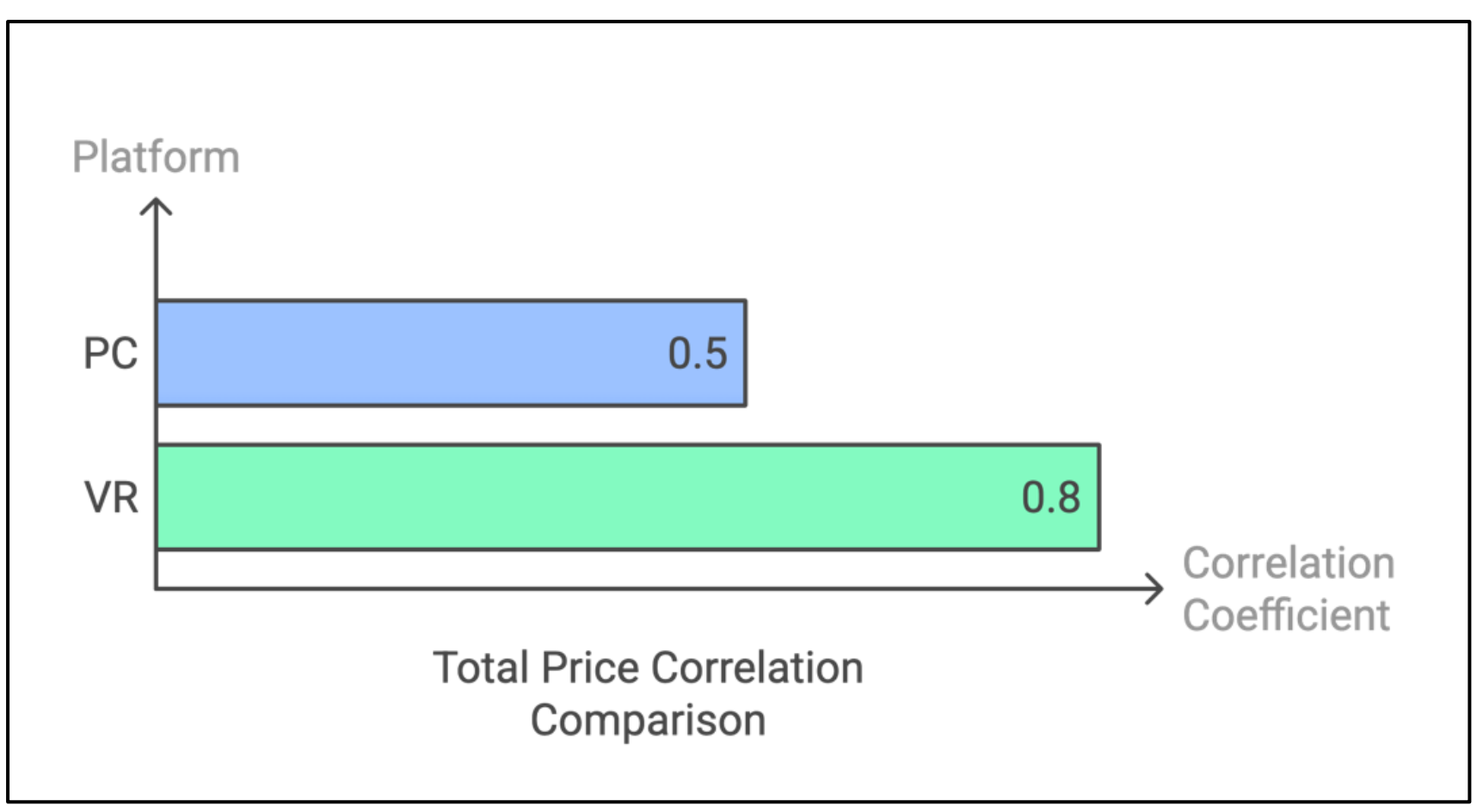

**Fig. 18** Total correlation between price category and high ratings: PC vs VR comparison

### 4.3 Comparing player enjoyment across VR and PC games

Table 14 presents the percentage of PC and VR games that received high ratings (4+ out of 5) across various game design elements. The overall average rating is higher for VR games (49.03%) than for PC games (33.82%), indicating that VR games generally receive more favorable reviews across most design aspects. However, the degree of enjoyment varies depending on the specific elements of game design.

**Table 14** Percentage of PC and VR games with high ratings for each game design element. Darker blue shading indicates higher percentages within each column

| Design Element | PC Games | VR Games |
|---|---|---|
| Graphics | 62.51 | 73.11 |
| Gameplay | 45.42 | 83.08 |
| Spatial Presence | 35.54 | 83.93 |
| Audio | 61.81 | 54.73 |
| Replayability | 27.06 | 65.46 |
| Monetization Model | 37.3 | 49.14 |
| Avatar Customization | 36.48 | 35.65 |
| Story | 36.7 | 33.85 |
| Community | 25.49 | 34.97 |
| Controls | 18.5 | 36.43 |
| Multiplayer | 14.65 | 33.51 |
| Difficulty | 4.39 | 4.55 |
| **Total Average** | 33.82% | 49.03% |

Among the individual design elements, 'Graphics' stand out as one of the highest-rated features on both platforms. A higher proportion of VR games (73.11%) receive strong ratings in this category compared to PC games (62.51%), reflecting a 10.6% difference in favor of VR. In contrast, audio design follows a different trend. While PC games (61.81%) receive a relatively high percentage of positive ratings, VR games (54.73%) fall slightly behind, marking the only category where PC games outperform VR.

Gameplay mechanics show the most pronounced disparity between the two platforms. VR games (83.08%) receive overwhelmingly higher ratings than PC games (45.42%), representing a 37.66% difference—one of the largest gaps in the dataset. The largest difference is observed in spatial presence, where VR games (83.93%) significantly outperform PC games (35.54%), with a 48.39% gap—the most substantial observed. Similarly, VR games also hold a major advantage in replayability, with 65.46% of VR titles receiving high ratings compared to 27.06% of PC games, marking a 38.4% difference.

Storytelling elements show little variation between the two formats, with PC games (36.7%) and VR games (33.85%) receiving comparable ratings, showing only a 2.85% difference. Monetization models also show a higher percentage of positive ratings in VR, with 49.14% of VR games receiving high ratings compared to 37.3% of PC games, a difference of 11.84%. Similarly, the multiplayer experience is rated more favorably in VR, with 33.51% of VR games receiving high ratings compared to 14.65% of PC games, showing an 18.86% gap. Finally, difficulty ratings remain fairly consistent across both platforms, with VR games (4.55%) and PC games (4.39%) receiving similarly low percentages of high ratings, showing only a 0.16% difference.

### 4.4 Highest-rated game design elements per game genre

#### 4.4.1. PC games

This analysis highlights the key game design elements that matter most across different genres in both PC and VR platforms, offering valuable insights for developers, designers, and publishers to optimize game features and enhance player experience.

Table 15 presents the average ratings of various game design elements across different PC game genres. Each row represents a genre, while each column corresponds to a specific game design element. The values in the table indicate the average rating given to each element within that genre.

**Table 15** Average ratings of game design elements across PC Game Genres. The term (RPG) in the table refers to 'Role-Playing Game'. Darker blue shading indicates higher average rating (out of 5) within each column. Leftmost column marks the highest rating per genre/game design element pair, while the rightmost column shows the lowest.

| Genre | Graphics | Audio | Avatar_Customization | Gameplay | Spatial_Presence | Story | Community | Monetization_Model | Replayability | Difficulty | Multiplayer | Controls |
|---|---|---|---|---|---|---|---|---|---|---|---|---|
| Music | 3.84 | 4.01 | 3.44 | 3.48 | 3.45 | 3.4 | 3.15 | 3.09 | 3.06 | 3 | 2.69 | 2.97 |
| Entertainment | 3.53 | 3.55 | 3.83 | 3.53 | 3.54 | 3.4 | 2.25 | 2.33 | 3.54 | 3.43 | 3.33 | 3 |
| Puzzle | 3.72 | 3.64 | 3.44 | 3.49 | 3.37 | 3.35 | 3.16 | 3.3 | 2.97 | 2.97 | 2.97 | 2.81 |
| Horror | 3.65 | 3.65 | 3.39 | 3.43 | 3.45 | 3.39 | 3.03 | 3.22 | 2.94 | 2.96 | 2.8 | 2.74 |
| Adventure | 3.65 | 3.62 | 3.44 | 3.42 | 3.32 | 3.29 | 3.15 | 3.11 | 2.99 | 2.96 | 2.91 | 2.75 |
| Education | 3.44 | 3.31 | 3.46 | 3.5 | 3.18 | 3.27 | 3.42 | 3.05 | 3.22 | 3.02 | 3.11 | 2.84 |
| Casual | 3.63 | 3.57 | 3.46 | 3.43 | 3.27 | 3.26 | 3.18 | 3.13 | 3.01 | 2.96 | 2.96 | 2.78 |
| Shooter | 3.58 | 3.57 | 3.45 | 3.51 | 3.37 | 3.13 | 3.02 | 3.04 | 3.09 | 2.93 | 2.83 | 2.99 |
| Action | 3.6 | 3.59 | 3.44 | 3.46 | 3.3 | 3.17 | 3.07 | 3.07 | 3.05 | 2.94 | 2.91 | 2.84 |
| Battle_Royale | 3.54 | 3.51 | 3.43 | 3.46 | 3.3 | 3.14 | 3.05 | 2.93 | 3.14 | 2.95 | 2.97 | 2.85 |
| RPG | 3.57 | 3.54 | 3.48 | 3.37 | 3.25 | 3.26 | 3.12 | 2.86 | 3.06 | 2.92 | 2.89 | 2.68 |
| Fighting | 3.57 | 3.58 | 3.39 | 3.52 | 3.22 | 3.2 | 2.8 | 2.74 | 3.15 | 2.94 | 2.95 | 2.87 |
| Strategy | 3.5 | 3.42 | 3.45 | 3.41 | 3.21 | 3.19 | 3.14 | 2.91 | 3.17 | 2.95 | 2.91 | 2.7 |
| Sports | 3.41 | 3.34 | 3.28 | 3.46 | 3.2 | 3.12 | 3.1 | 2.98 | 3.23 | 2.96 | 2.97 | 2.89 |
| Racing | 3.43 | 3.33 | 3.4 | 3.45 | 3.22 | 3.08 | 2.97 | 3.1 | 3.15 | 2.97 | 2.9 | 2.88 |
| Simulation | 3.51 | 3.43 | 3.45 | 3.34 | 3.24 | 3.22 | 3.21 | 2.92 | 3.03 | 2.95 | 2.85 | 2.61 |
| Survival | 3.5 | 3.49 | 3.42 | 3.32 | 3.27 | 3.15 | 3.19 | 2.91 | 2.98 | 2.92 | 2.79 | 2.56 |

'Graphics' and 'Audio' tend to have relatively high scores across all genres, with 'Music' (3.84) and 'Puzzle' (3.72). While 'Audio' receives the highest scores in 'Music' (4.01) and 'Entertainment' (3.55). 'Avatar Customization' is consistently rated between 3.3 and 3.8 across genres, with the highest value found in 'Entertainment' (3.83). 'Gameplay' scores are relatively consistent across genres, with 'Entertainment' (3.53) and 'Strategy' (3.41) at the higher end, while 'Spatial Presence' scores range between 3.1 and 3.4, with 'Shooter' (3.37) and 'Horror' (3.45) being among the highest-rated.

'Story' ratings show variation across genres, with 'Horror' (3.39), 'RPG' (3.26), and 'Adventure' (3.29) having the highest values, whereas genres like 'Shooter' (3.13) and 'Sports' (3.12) have lower ratings. 'Community' ratings are in the 3.0 to 3.4 range, with 'Education' (3.42) and 'Adventure' (3.15) scoring higher, while 'Entertainment' (2.25) has the lowest rating.

'Monetization Model' generally has the lowest scores among all elements, with 'Fighting' (2.74) and 'Strategy' (2.91) being at the lower end, while 'Puzzle' (3.30) and 'Casual' (3.13) score slightly higher. 'Replayability' ratings vary, with 'Entertainment' (3.54) having the highest value, while genres like 'Horror' (2.94) and 'Puzzle' (2.97) have lower scores. 'Difficulty' ratings are relatively close across genres, with 'Entertainment' (3.43) and 'Music' (3.00) at the higher end, and 'Strategy' (2.95) and 'Simulation' (2.95) at the lower end.

'Multiplayer' ratings are highest in 'Sports' (2.98) and 'Fighting' (2.95), while 'Music' (2.69) and 'Horror' (2.80) rate it the lowest. 'Controls' ratings fluctuate, with 'Strategy' (3.17) and 'Racing' (3.15) having the highest values, and 'Survival' (2.56) and 'Horror' (2.74) ranking lower.

#### 4.4.2. VR games

Like Table 15, Table 16 presents the average ratings of various game design elements across VR game genres instead. Unlike in PC games, ‘Spatial Presence’ scores highly across all VR genres, with ‘Strategy’ (4.23), ‘Entertainment’ (4.33), and ‘Shooter’ (4.06) receiving the highest ratings, while even the lowest-scoring genres, ‘Education’ (3.94) and ‘Battle Royale’ (3.92), remain relatively high. This contrasts with PC games, where ‘Spatial Presence’ varied more significantly across genres, showing that immersion is a more universal expectation in VR. ‘Gameplay’ ratings remain strong across genres, similar to PC, with ‘Entertainment’ (4.00), ‘Sports’ (3.91), and ‘Strategy’ (3.85) leading, while games tagged with ‘Adventure’ (3.76) and ‘Role-Playing Game’ (3.71) have slightly lower ratings. In VR, the spread of ‘Gameplay’ ratings is narrower than in PC games, where RPGs showed more differentiation.

**Table 16** Average ratings of game design elements across VR game genres. The term (RPG) in the table refers to ’Role-Playing Game’. Darker blue shading indicates higher average rating (out of 5) within each column. Leftmost column marks the highest rating per genre/game design element pair, while the rightmost column shows the lowest.

| Genre | Spatial_Presence | Gameplay | Graphics | Replayability | Audio | Monetization_Model | Avatar_Customization | Community | Story | Multiplayer | Controls | Difficulty |
|---|---|---|---|---|---|---|---|---|---|---|---|---|
| Music | 4.04 | 3.86 | 3.61 | 3.82 | 3.98 | 3.41 | 3.52 | 3.56 | 3.26 | 3.13 | 3.42 | 3.05 |
| Puzzle | 4.08 | 3.77 | 3.9 | 3.44 | 3.65 | 3.63 | 3.4 | 3.44 | 3.46 | 3.28 | 3.14 | 3.01 |
| Sports | 4.05 | 3.91 | 3.65 | 3.85 | 3.34 | 3.46 | 3.56 | 3.37 | 3.26 | 3.33 | 3.24 | 3 |
| Strategy | 4.23 | 3.85 | 3.86 | 3.67 | 3.51 | 3.4 | 3.36 | 3.24 | 3.33 | 3.15 | 3.35 | 3 |
| Casual | 3.99 | 3.79 | 3.75 | 3.52 | 3.59 | 3.49 | 3.39 | 3.46 | 3.41 | 3.36 | 3.15 | 3.04 |
| Shooter | 4.06 | 3.83 | 3.69 | 3.68 | 3.45 | 3.63 | 3.42 | 3.36 | 3.31 | 3.19 | 3.28 | 3.02 |
| Action | 4.02 | 3.86 | 3.72 | 3.72 | 3.53 | 3.51 | 3.43 | 3.34 | 3.36 | 3.19 | 3.23 | 2.99 |
| Education | 3.94 | 3.71 | 3.76 | 3.31 | 3.47 | 3.71 | 3.71 | 3.85 | 3.29 | 3.58 | 2.76 | 3.06 |
| Adventure | 4.03 | 3.76 | 3.79 | 3.46 | 3.6 | 3.48 | 3.41 | 3.4 | 3.48 | 3.21 | 3.08 | 3 |
| Survival | 4.03 | 3.86 | 3.63 | 3.72 | 3.47 | 3.54 | 3.44 | 3.52 | 3.36 | 3.16 | 3.05 | 2.99 |
| Entertainment | 4.33 | 4 | 4 | 2.5 | 2.67 | 3.67 | 4 | 3.5 | 4 | 3 | 3 | 3 |
| Fighting | 4.02 | 3.82 | 3.76 | 3.64 | 3.35 | 3.5 | 3.51 | 3.37 | 3.28 | 3.16 | 3.26 | 2.97 |
| RPG | 3.94 | 3.71 | 3.66 | 3.64 | 3.52 | 3.3 | 3.38 | 3.67 | 3.25 | 3.29 | 3.14 | 3 |
| Racing | 3.91 | 3.85 | 3.7 | 3.67 | 3.39 | 3.64 | 3.38 | 3.22 | 3.24 | 3.09 | 3.28 | 3 |
| Simulation | 3.99 | 3.81 | 3.61 | 3.67 | 3.37 | 3.4 | 3.46 | 3.49 | 3.27 | 3.12 | 3.09 | 3.02 |
| Battle_Royale | 3.92 | 3.68 | 3.57 | 3.44 | 3.43 | 3.39 | 3.44 | 3.35 | 3.23 | 3.36 | 3.02 | 3.03 |
| Horror | 3.92 | 3.77 | 3.64 | 3.44 | 3.48 | 3.36 | 3.44 | 3.36 | 3.36 | 3.18 | 2.84 | 2.95 |

‘Graphics’ ratings are consistently high across all genres, with ‘Entertainment’ (4.00), ‘Strategy’ (3.86), and ‘Puzzle’ (3.90) leading, while ‘Battle Royale’ (3.57) and ‘Horror’ (3.64) are among the lower-rated. The highest PC game ratings for ‘Graphics’ were around 3.84, meaning VR games tend to have slightly higher visual ratings overall, likely due to their dependence on high-quality visuals for immersion.

‘Replayability’ in VR shows a different distribution compared to PC, with ‘Sports’ (3.85), ‘Music’ (3.82), and ‘Shooter’ (3.68) being the highest-rated, while ‘Entertainment’ (2.50) is significantly lower than in PC. This suggests that VR entertainment applications often provide one-time immersive experiences rather than repeatable ‘Gameplay’ sessions.

‘Audio’ ratings in VR are highest in ‘Music’ (3.98), ‘Adventure’ (3.60), and ‘Education’ (3.47), whereas in PC, ‘Shooter’ (3.57) and ‘Fighting’ (3.58) had stronger ‘Audio’ ratings. While VR ratings remain more concentrated, PC games had a broader range of ‘Audio’ scores, showing that audio quality is prioritized differently between the two platforms.

‘Monetization Model’ ratings remain among the lowest-rated elements in both VR and PC. The highest VR scores appear in ‘Puzzle’ (3.63), ‘Shooter’ (3.63), and ‘Sports’ (3.46), while ‘Strategy’ (3.40) and ‘Fighting’ (3.50) rank lower, a pattern similar to PC games, where monetization was one of the least favored aspects across most genres.

‘Avatar Customization’ is rated highly in ‘Entertainment’ (4.00), ‘Sports’ (3.56), and ‘Education’ (3.71), while ‘Fighting’ (3.51) and ‘Battle Royale’ (3.43) are at the lower end. In PC games, ‘Avatar Customization’ followed a similar distribution, though the difference between highest- and lowest-rated genres is more pronounced in VR.

‘Community’ ratings in VR are strongest in ‘Education’ (3.85) and ‘Entertainment’ (3.50), while ‘Fighting’ (3.37) and ‘Shooter’ (3.36) have lower ratings. This follows the same pattern as PC games, where ‘Education’ and ‘Entertainment’ showed higher ‘Community’ engagement compared to competitive genres.

‘Story’ ratings are highest in ‘Entertainment’ (4.00), ‘Role-Playing Game’ (3.48), and ‘Strategy’ (3.33), while ‘Sports’ (3.26) and ‘Fighting’ (3.28) score lower. While ‘Role-Playing Games’ and ‘Adventure’ games had similarly high ‘Story’ ratings in both VR and PC, ‘Entertainment’ games score significantly higher in VR, suggesting a stronger focus on immersive storytelling.

‘Multiplayer’ ratings in VR are strongest in ‘Role-Playing Game’ (3.29), ‘Adventure’ (3.21), and ‘Sports’ (3.33), while ‘Fighting’ (3.16) and ‘Shooter’ (3.19) score slightly lower. Compared to PC, ‘Multiplayer’ scores in ‘Role-Playing Games’ are higher in VR, suggesting that social VR role-playing experiences are more emphasized than in traditional PC games.

‘Controls’ ratings are highest in ‘Shooter’ (3.28), ‘Strategy’ (3.35), and ‘Racing’ (3.28), while ‘Education’ (2.76) and ‘Horror’ (2.84) have the lowest ratings. PC games also showed high ‘Controls’ ratings for ‘Strategy’ and ‘Racing’, though ‘Education’ did not rank as low as it does in VR.

‘Difficulty’ ratings in VR are relatively even across all genres, with ‘Education’ (3.06) ranking highest, while ‘Fighting’ (2.97) and ‘Horror’ (2.95) rank lowest. This differs from PC games, where ‘Fighting’ had higher ‘Difficulty’ ratings, suggesting that VR developers may prioritize accessibility over challenge in these genres.

Figures 19 and 20 illustrate the standard deviations of the averages presented in Tables 15 and 16. The results indicate that variation in ratings across genres is generally small, with values ranging from 0.06 to 0.51. This suggests that most elements are rated fairly consistently across different genres. For example, ‘Difficulty’ (0.06) shows almost no variation, meaning players perceive challenge levels similarly regardless of genre on both PC and VR. In contrast, ‘Multiplayer’ (0.51) exhibits the widest spread on PC, making it highly genre-dependent.

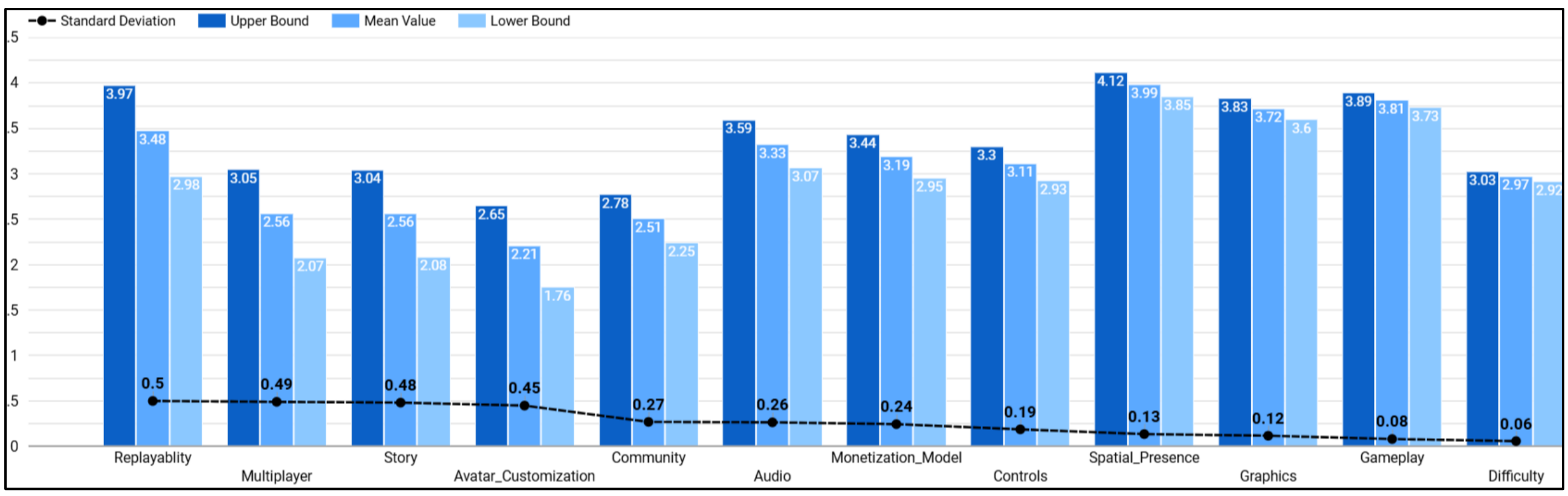

**Fig. 19** Standard deviation for VR average ratings. Design elements are ordered left-to-right by decreasing standard deviation (highest to lowest)

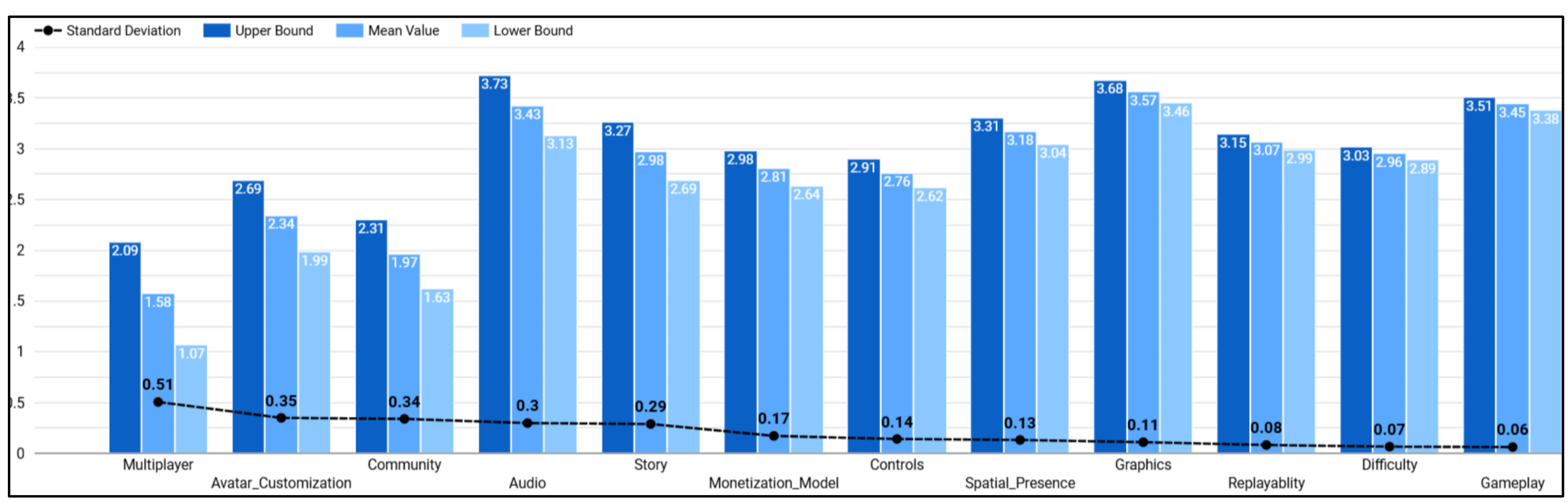

**Fig. 20** Standard deviation for PC average ratings. Sorted by standard deviation. Design elements are ordered left-to-right by decreasing standard deviation (highest to lowest)

In comparison, differences across platforms are more moderate but still meaningful. As shown in Table 17, values range from as low as 0.01 up to about 0.70, indicating that certain ‘Design Elements’ vary more noticeably between PC and VR experiences. For instance, ‘Multiplayer’ (0.70) and ‘Spatial Presence’ (0.57) show the strongest platform differences, suggesting that social interaction and immersive presence are perceived quite differently depending on the medium (PC or VR). Conversely, 'Difficulty' (0.01) and 'Audio' (0.07) remain highly consistent, with little perceptual difference between PC and VR players.

**Table 17.** Standard deviation of design elements across PC and VR platforms

| Design Element | Value |
|---|---|
| Multiplayer | 0.7 |
| Spatial_Presence | 0.57 |
| Community | 0.39 |
| Replayability | 0.29 |
| Story | 0.29 |
| Monetization_Model | 0.27 |
| Gameplay | 0.26 |
| Controls | 0.25 |
| Graphics | 0.11 |
| Avatar_Customization | 0.1 |
| Audio | 0.07 |
| Difficulty | 0.01 |

### 4.5 Predicting what game genres have the biggest impact on ratings

Table 18 presents the predicted average overall rating for different game genres in VR and PC gaming, based on the XGBoost regression model. The values represent the expected user enjoyment scores for each genre on their respective platforms.

**Table 18** Predicted average ratings of game genres in VR vs PC

**Prediction - Genre Influence on VR vs PC Ratings**

| VR Games | | PC Games | |
|---|---|---|---|
| Genre | Predicted Average Overall Rating | Genre | Predicted Average Overall Rating |
| Role-Playing Game | 3.14 | Battle Royale | 2.89 |
| Strategy | 3.14 | Fighting | 2.88 |
| Battle Royale | 3.13 | Sports | 2.87 |
| Music | 3.09 | Music | 2.86 |
| Shooter | 3.09 | Shooter | 2.85 |
| Fighting | 3.09 | Action | 2.85 |
| Sports | 3.08 | Racing | 2.84 |
| Survival | 3.08 | Strategy | 2.82 |
| Racing | 3.07 | Role-Playing Game | 2.81 |
| Simulation | 3.06 | Casual | 2.79 |
| Action | 3.04 | Adventure | 2.78 |
| Casual | 3.04 | Simulation | 2.78 |
| Adventure | 3.02 | Survival | 2.77 |
| Puzzle | 3.01 | Horror | 2.75 |
| Horror | 2.97 | Puzzle | 2.73 |
| Education | 2.95 | Education | 2.73 |
| Entertainment | 2.36 | Entertainment | 2.64 |

In the VR Games category, the highest predicted ratings are for 'Role-Playing Game' (3.14) and 'Strategy' (3.14), both receiving the same expected enjoyment level. 'Battle Royale' (3.13) follows closely behind, with only a minor difference from the top-rated genres. 'Music' (3.09), 'Shooter' (3.09), and 'Fighting' (3.09) share identical predicted ratings, forming the next group of highly-rated genres, slightly below the leading ones. 'Sports' (3.08) and 'Survival' (3.08) also have nearly identical predicted values, ranking just below the previous group. 'Racing' (3.07) and 'Simulation' (3.06) follow closely, maintaining a small gap from the higher-scoring genres. Further down, 'Action' (3.04), 'Casual' (3.04), and 'Adventure' (3.02) show slight decreases in predicted ratings, with 'Puzzle' (3.01) positioned just below them. ‘Horror’ (2.97) ranks slightly lower than ‘Puzzle’, while ‘Education’ (2.95) follows closely. The lowest-rated genre in this category is ‘Entertainment’ (2.36), which is significantly lower than all other genres.

While ‘Role-Playing Game’ and ‘Strategy’ rank at the top in VR gaming, ‘Battle Royale’ (2.89) receives the highest predicted rating in PC games. ‘Fighting’ (2.88), ‘Sports’ (2.87), and ‘Music’ (2.86) follow closely, all within a narrow range. ‘Shooter’ (2.85) and ‘Action’ (2.85) maintain identical predicted scores, positioning them just below the leading group. Further down, ‘Racing’ (2.84), ‘Strategy’ (2.82), and ‘Role-Playing Game’ (2.81) show small differences, indicating a slightly lower expected enjoyment compared to their VR counterparts (3.14 for both ‘Role-Playing Game’ and ‘Strategy’). ‘Casual’ (2.79), ‘Adventure’ (2.78), and ‘Simulation’ (2.78) are positioned closely together, showing minimal variation in their expected ratings. ‘Survival’ (2.77) and ‘Horror’ (2.75) rank just below them, while ‘Puzzle’ (2.73) is next in line. At the lower end of the predictions, ‘Education’ (2.73) and ‘Entertainment’ (2.64) receive the lowest ratings, with ‘Entertainment’ ranking the lowest among all PC game genres. While ‘Entertainment’ games score slightly higher in PC gaming (2.64) compared to VR (2.36), they remain the least favorably rated genre overall.

# 5 Discussion

## 5.1 Discussion of key findings

### 5.1.1. Research question 1: What are the key game design features that contribute most significantly to player enjoyment in PC versus VR games?

RQ1 examines which design features matter most on each platform. Our results point to ‘Spatial Presence’, Graphics’, ‘Audio’, and ‘Gameplay’ as the most consistent drivers of high ratings, with clear platform specific patterns. ‘Spatial Presence’ stands out in VR, where immersion is valued more strongly than on PC. This supports the common view that presence is central to virtual reality [46]. Audio relates closely to overall reception on both platforms, though PC titles tend to edge VR slightly. This may reflect higher expectations for spatial and three-dimensional sound in immersive settings. ‘Gameplay’ and ‘Replayability’ also favor VR. Players often describe the physical, hands-on interaction as fresh and engaging across sessions. This kind of deeply engaging, interactive play is reminiscent of a flow state, a condition of optimal experience where challenge and skill are balanced, which can greatly enhance enjoyment [46]. By contrast, on PC sustained interest more often comes from broad content variety, social features, or community activity. The emphasis on community and social play in PC titles suggests that players value social connectedness, aligning with the relatedness need described in self-determination theory [47].

### 5.1.2. Research question 2: In what ways do monetization strategies affect player enjoyment in free-to-play versus non-free-to-play games?

RQ2 looks at how monetization shapes player enjoyment across free to play and non free to play titles. In PC games, pricey releases that add microtransactions or frequent DLC tend to draw strong pushback, which is in line with prior work on player resistance to paywalls [48]. From a motivation perspective, this reaction reflects how aggressive monetization can undermine players’ sense of autonomy and intrinsic motivation, thereby diminishing enjoyment [47]. Players generally want to feel in control of their play experience; if monetization schemes (e.g., paywalls or pay-to-win mechanics) are perceived as controlling or unfair, they erode the fun. Indeed, self-determination theory posits that when external rewards or costs pressure the player, it can thwart autonomy and reduce intrinsic enjoyment [47]. Multiplayer reception also slips as price rises, while lower cost and free to play multiplayer games often build stronger communities, as seen with League of Legends and CS,GO [49], [50]. The thriving communities in these free titles suggest that social bonds can compensate for or even enhance enjoyment in the absence of upfront costs. In other words, a vibrant community can fulfill players’ relatedness needs, reinforcing their enjoyment even when the game monetization relies on optional purchases [47].

Furthermore, price relates to several design elements. Higher priced games more often focus on avatar customization, graphics, controls, and replayability, which fits the idea that premium titles invest in visuals, polish, player expression, and long term value [51]. By contrast, cheaper PC games tend to feel more demanding, which suggests sharper learning curves or less refined mechanics, while premium titles aim for a more accessible experience [52].

Story shows a split by platform. In VR, non free to play games are closely linked with richer storytelling, hinting that premium VR releases put more weight on narrative quality. In PC, story quality appears less tied to price, which means strong narratives can show up at many price points.

On the other hand, VR pricing shows only a weak link with how players view monetization [53]. Many VR players accept paying up front and have seen fewer pay to win or microtransaction heavy models. VR also receives a larger share of positive views on monetization compared to PC. Taken together, if a PC game is free to play, players are more open to cosmetic or optional purchases, while premium PC games that stack on extra costs are judged more harshly.

#### 5.1.3. Research question 3: What differences exist in player enjoyment between PC and VR game genres?

RQ3 examines how enjoyment differs by genre across PC and VR. Overall, VR titles tend to receive warmer responses, and this is most visible in features such as 'Gameplay', 'Spatial Presence', and 'Replayability'. These patterns agree with prior work on VR user experience that links immersion with higher enjoyment [54].

By genre, games tagged with 'Role-Playing Game' and 'Strategy' stand out in VR, where immersion and deliberate decision making fit the strengths of the medium. These trends suggest that the strong presence and embodied play in VR can elevate genres reliant on narrative, tactics, or exploration. For example, a rich RPG story or complex strategy scenario may feel more compelling when the player is immersed in it [46]. 'Battle Royale' titles perform best on PC yet still earn solid reception in VR, indicating that fast-paced competitive play is enjoyable on both platforms, though players might appreciate it for different reasons in each context.

Clear platform preferences also appear. PC games lean toward 'Casual' and 'Adventure' genres, reflecting broad reach and easy entry. VR catalogs favor 'Action' and 'Simulation' genres, where hands on interaction and spatial awareness are central. Multiplayer experiences often feel more engaging in VR, helped by the heightened sense of presence, spatial audio, and the ability to read others' body language in-game. This richer social presence in VR means cooperative or competitive play can better satisfy players' social needs, an echo of the relatedness factor in player motivation theories detailed by Grasse and his team [47]. In an RPG, for instance, interacting with other players' avatars in an embodied way can deepen camaraderie and enjoyment. 'Controls' and precision matter in both ecosystems (particularly for genres like 'Racing' or 'Strategy'), but we see that VR releases often aim for approachability over very high difficulty in genres such as 'Fighting' or 'Horror'. This difference might be due to VR developers balancing challenge to avoid player frustration or motion discomfort. From a flow perspective, keeping difficulty approachable helps a wider range of players enter a flow state of enjoyment rather than becoming overwhelmed [46]. In summary, the genre-based findings reinforce that VR's strength lies in providing immersive, physically engaging experiences that align with theories of presence and flow, whereas PC's strength is its versatility and social connectivity, speaking to a wide spectrum of player motivations.

#### 5.1.4. Research question 4: How can generative AI techniques be employed to identify the primary factors influencing positive player reviews across different game genres and platforms?

RQ4 looks at how generative AI can surface what drives positive reviews across genres and platforms. Our pipeline uses large scale automated text summarization, sentiment analysis, and topic modeling. Together, these methods reveal patterns that are hard to find by hand.

Sentiment from player comments helps point to problem pairs, for example price with difficulty, and price with monetization, and it flags areas that may need attention, such as VR audio design. Topic modeling groups comments into clear themes. These include pay to win frustration, concerns about fairness in skill based systems, and dislike of short campaigns in expensive VR adventure games. Prior work, for example Guzsvinecz, shows that these methods can also expose genre specific details that do not always appear with manual coding [5].

In sum, generative AI offers a scalable and data driven way to learn which design features attract players and which ones push them away. This lets developers and researchers update their games and studies with more focus and better precision.

### 5.2 Implications of the study

The findings from this study offer practical and theoretical implications that can guide both industry practitioners and academic researchers. In this section we focus on how to act on them.

First, platform matters. Developers should tailor design to the strengths of each platform. In VR, prioritize immersion and physical engagement, with strong motion controls, high fidelity visuals, and polished interactive mechanics that support 'Spatial Presence', 'Gameplay', and 'Replayability'. On PC, give careful attention to audio design and to monetization choices. Players respond better when pricing and added purchases feel transparent and fair, which helps sustain goodwill and loyalty.

Second, integrate generative AI into the development and research cycle. Automated text summarization, sentiment analysis, and topic modeling let teams process large volumes of feedback, surface recurring pain points, and monitor how attitudes change over time. This supports quicker design pivots and clearer product decisions. The same approach can extend to mobile and console as cross-platform play grows.

Third, match genre to platform strengths. VR tends to reward genres that lean on deep interaction and narrative, such as RPG and Strategy, while some casual or educational titles may struggle to reach high enjoyment without careful adaptation. Aligning content with what each platform does best is essential for positive user experiences.

### 5.3 Limitations of the study

While this study provides valuable insights into player sentiment and game design trends in the post pandemic gaming landscape, several limitations must be acknowledged. The focus on games released between 2020 and 2024, while justifiable due to shifts in player behavior, restricts the generalizability of findings to earlier or future releases. The presence of review bombing also poses a challenge for sentiment analysis, which may not always reflect genuine player experiences [7].

In addition, relying on generative AI for text tokenization can introduce bias, since models may misread sarcasm, humor, or context, which can affect sentiment classification [12]. Finally, due to limited resources and the compute required to process each review with the Phi 4 SLM, only the 100 most rated reviews were analyzed for each of the 4,856 PC and VR games. This sampling choice can underrepresent lesser known or niche titles and can emphasize mainstream preferences, which may overlook perspectives from smaller player communities.

## 6 Conclusion and future works

In this study, we analyzed reviews from 4,856 PC and VR games, using a generative AI pipeline to uncover what drives player enjoyment across genres and platforms. Below are the takeaways:

- Platform matters. VR tends to earn stronger responses for immersion and hands on interaction, while PC strengths vary by element and audience. Designing to each platform's capabilities is key.
- Monetization shapes sentiment, especially for premium PC titles where added costs can strain goodwill. Clear and optional purchase models are received more positively. VR players are generally more comfortable with paying up front.
- Genre and platform should be matched with care. Story heavy and strategy focused experiences align well with VR's sense of presence, while several competitive and pick up genres remain a better fit for PC.
- Generative AI methods, including text summarization and topic modeling, help teams process large volumes of feedback and surface themes such as concerns about pay to win and the importance of fair difficulty. These tools provide a practical, data driven basis for design decisions.

Together, these points suggest that understanding the interplay among immersion, monetization, and player expectations is central to making more engaging games. Developers and publishers can use these insights to tune price models, adjust difficulty, and refine interactive systems. Researchers can apply the same approach to other platforms and contexts to test how robust these patterns are.

Future studies should widen the release window beyond 2020 to 2024 so results generalize to earlier and future titles. Methods that detect and reduce review bombing are also needed, for example agent-based AI workflows that combine gameplay telemetry, community discussions, and streaming interactions to spot manipulative or inauthentic reviews. Model accuracy can improve by refining Transformer models, so they read sarcasm, humor, and context more reliably, supported by targeted human checks where needed. Finally, to reduce sampling bias, researchers can analyze more than the top 100 reviews per game and use stratified sampling that includes both popular and niche titles, while investing in more efficient processing so larger review sets are practical.

## Ethical approval

This study was conducted in accordance with the ethical guidelines set forth by the University of Wollongong in Dubai. Given that this research involved the analysis of publicly available game reviews, it did not require formal approval from an institutional ethics review board. No personally identifiable data were collected, and all user-generated content was anonymized in adherence to ethical research principles.

## Consent to participate

Not applicable. This study utilized publicly available data from digital game distribution platforms (Steam and Meta Quest) without direct interaction with human participants. As such, no consent to participate was required.

## Consent to publish

Not applicable. The dataset used in this study consists of publicly accessible game reviews, and no proprietary or confidential information was disclosed. The findings of this research are presented in aggregate form to ensure no individual user can be identified.

## Data availability statement

The datasets analyzed in this study were collected from publicly accessible sources, namely Steam and the Meta Quest stores. The processed dataset, including quantified game design elements and metadata, is available upon request. Code used for data processing and analysis is available upon request.

## Authors’ contributions

Hisham Abdelqader conceptualized the study, designed the methodology, and carried out the data collection and processing. The data analysis and visualization were conducted by the author using generative AI and machine learning techniques. The manuscript was drafted and revised by Hisham Abdelqader. The author takes full responsibility for the integrity and accuracy of the study’s findings.

## Funding

This research received no specific grant from any funding agency, commercial entity, or not-for-profit organization. All computational resources were self-funded, and the study was conducted independently without external financial support.

## Competing interests

The author declares no competing interests. The study was carried out independently, and no conflicts of interest, financial or otherwise, influenced the research process or its findings.